\documentclass[%
 reprint,
superscriptaddress,
 amsmath,amssymb,
 aps,
]{revtex4-2}

\usepackage{graphicx}
\usepackage{dcolumn}
\usepackage{bm}
\usepackage[T1]{fontenc}

\begin{document}

\preprint{APS/123-QED}

\title{Spin-dependent routing of optical beams in the bulk of twisted anisotropic media}

\author{Chandroth P. Jisha}
 \email{jisha.chandroth.pannian@uni-jena.de}
 \affiliation{%
    Friedrich Schiller University Jena, Institute of Applied Physics, Abbe Center of Photonics,  Albert-Einstein-Str. 15, 07745 Jena, Germany
}%

\author{Lorenzo Marrucci}

\affiliation{
 Dipartimento di Fisica ``Ettore Pancini'', Universit\`{a} di Napoli Federico II, Complesso Universitario di Monte Sant’Angelo, Via Cintia, 80126 Napoli, Italy
}%

\author{Stefan Nolte}%
\affiliation{%
   Friedrich Schiller University Jena, Institute of Applied Physics, Abbe Center of Photonics,  Albert-Einstein-Str. 15, 07745 Jena, Germany
}%
\affiliation{Fraunhofer Institute for Applied Optics and Precision Engineering IOF, Albert-Einstein-Str. 7, 07745 Jena, Germany}

\author{Alessandro Alberucci}%
\email{alessandro.alberucci@uni-jena.de}
\affiliation{%
   Friedrich Schiller University Jena, Institute of Applied Physics, Abbe Center of Photonics, Albert-Einstein-Str. 15, 07745 Jena, Germany
}%
\date{\today}

\begin{abstract}
We theoretically discuss a new kind of photonic spin-Hall effect (PSHE) for optical beams propagating inside an inhomogeneously twisted anisotropic material. The rotation angle plays the role of an effective gauge field. When the twisting distribution is odd symmetric, the optical beams move along mirror-symmetric trajectories according to their helicity. Connection of this volumetric PSHE with the geometric phase  and the Kapitza effect is elucidated.  
\end{abstract}

\maketitle


\section{Introduction: Analogous of magnetic effects in optics}

Magnetism is one of the most fascinating class of phenomena in physics, with several deep implications concerning both fundamental phenomena and applications. For the purposes of the current paper, we want to focus on two specific effects, the Stern-Gerlach (SG) experiment \cite{Gerlach:1922} \footnote{Hereafter, we will use the term Stern-Gerlach effect to indicate the symmetric spatial separation between beams carrying opposite spins or pseudo-spins.} and the spin-Hall effect (SHE) \cite{Hirsch:1999}. In their historic experiment, Stern and Gerlach proved how the deflection of magnetic atoms in an inhomogeneous magnetic field produces two main lobes. Given that the magnetic energy can be cast in the Zeeman form  $\bm{\mu}\cdot \bf B$, such an experiment shows that only two discrete values are allowed for the spin magnetic moment $\bm \mu$. The SHE is instead the generation of a spin current once a charge current is flowing in a material. Such a current is due to a spin-dependent deflection, stemming from the spin-orbit interaction of the material \cite{Sinova:2015}. Although both the effects consists of a spin-dependent deflection, there is an essential difference to discriminate between them: the SHE and the SG effect come from an inhomogeneity in the magnetic field in the reciprocal and real space, respectively \cite{Lu:2006}. Interestingly, a beam of atoms subject to a gradient in optical intensity has been used to demonstrate experimentally an all-optical SG effect \cite{Sleator:1996}. Another all-optical case has been recently demonstrated in parametric nonlinear crystals subject to a spatial gradient in their nonlinear properties \cite{Yesharim:2022}.

On the other side, due to relative ease of optical experiments with respect to other fields, optics have been widely used to investigate general phenomena \cite{Longhi:2009,Patsyk:2020,Marques:2023}, or even like a workbench for analogues of effects which cannot be empirically tested in a controllable manner in other physical fields \cite{Paredes:2020}. Magnetic phenomena are not excluded from this list, including the observation of topological insulators \cite{Rechtsman:2013,Klembt:2018}, the observation of the Landau levels \cite{Schine:2016}, the direct observation of the Laughlin states \cite{Clark:2020}, and so on. Actually, a natural connection between light propagation in anisotropic materials and evolution of spin 1/2 particles under the influence of a magnetic field has been explicitly outlined \cite{Kuratsuj:1998}. Photonic SG deflection and Photonics SHE (PSHE) are not an exception to this general trend: several different types of spin/polarization-dependent trajectories have been reported in literature \cite{Ling:2017,Liu:2017,Kim:2023,Yesharim:2022,Zhou:2024}. Spin-orbit interaction indeed became a field of extreme interest in optics \cite{Cardano:2015,Bliokh:2015}. 

A large number of the examples of PSHE are based upon gradients in the geometric phase, both in the real and in the reciprocal phase space \cite{Ling:2017}. In the first case, it is related with the rotation of the polarization for a fixed wavevector, and it is called Pancharatnam-Berry phase (PBP). In the second case, the geometric phase is due to the rotation of the wavevector and it is called spin-redirection phase (SRP).  SRP can for example induce a spin-dependent trajectory in inhomogeneous isotropic materials, also called as optical Magnus effect \cite{Onoda:2004,Bliokh:2004_1,Bliokh:2008_1}. Bliokh in 2007 demonstrated that SRP can explain the tiny polarization-dependent shifts occurring at optical interfaces \cite{Bliokh:2007_1}, already known in optics as the Imbert-Fedorov shift \cite{Bliokh:2013}. Such an effect has been later demonstrated borrowing weak measurement theory from quantum mechanics to optics \cite{Hosten:2008}. On the other hand, angular deflections due to the PBP have been introduced some years earlier using sub-wavelength spatially-varying gratings \cite{Hasman:2002}, to be then vastly investigated using liquid crystals \cite{Marrucci:2006_1}, laser-written nanostructures \cite{Drevinskas:2017_1} and metasurfaces \cite{Liu:2017}, for example showing a huge enhancement in the presence of a $\mathcal{PT}$-symmetric system \cite{Zhou:2019}. \\
Beyond these two most famous examples, optics provides a plethora of material responses and experimental settings where this kind of effects can be observed: even in free space, optical surface waves show a spin-momentum locking \cite{Bliokh:2015_1}. Spin-dependent coupling has been explored in plasmonic devices \cite{Gorodetski:2008,Rodriguez:2013}. PSHE can also be observed when a tight focused beam crosses a tilted but homogeneous uniaxial material \cite{Bliokh:2016}. Focusing on anisotropic materials, in a simple homogeneous crystal the extraordinary component in the general case does not overlap with the ordinary component due to the walk-off angle \cite{Yariv:1984}, an effect which can be described as induced by the presence of a homogeneous potential vector \cite{Alberucci:2010}. Generalizations have been explored along different directions: by considering the wavevector degrees of freedom, inhomogeneous magnetic effects in the reciprocal space have been shown in homogeneous crystals \cite{Ling:2020,Zhu:2021}; by using electro-optical modulation, either yielding a twisted \cite{Zhu:2024} or an inhomogeneous magnetic field proportional to the propagation constant mismatch \cite{Liu:2024}; by using materials encompassing a bi-anisotropic response, examples of non-trivial bidimensional potential vectors have been discussed \cite{Liu:2015}; by filling up optical cavities with liquid crystals to engineer spin-orbit Hamiltonians \cite{Rechcinska:2019}, yielding e.g. an analogue of SG effect \cite{Krol:2021}. Finally, PSHE due to nonlinear effects has been observed using reorientational nonlinearity in liquid crystals \cite{ElKetara:2013}, whereas the equivalent of a SG effect has been demonstrated in second harmonic generation in nonlinear structured crystals \cite{Yesharim:2022}.

The goal of the current paper is to demonstrate theoretically and numerically a new kind of spin-dependent trajectories in twisted anisotropic materials, the latter being homogeneous along the propagation direction. Unlike PSHE based upon thin (with respect to the Rayleigh distance of the beam) twisted samples matching the half-wave plate condition, the effect emerges in the bulk of the material as an interplay between the natural diffraction and a longitudinally-periodic gradient in the geometric phase. By rewriting the wave propagation in a reference system everywhere parallel to the local orientation of the crystal, we show that light evolves under the action of an effective gauge field proportional to the gradient of the rotation angle.  
With respect to PSHE occurring at abrupt interfaces, we will show that deflections much larger than the beam widths can be easily achieved, the maximum deflection being actually related to the overall length of the samples. 

The Paper is structured as follows. In Sec.~\ref{sec:basic_idea} we expose the basic idea and provide the basic model based upon a geometrical optics model. In Sec.~\ref{sec:FDTD} we confirm the existence of the PSHE using FDTD simulations. The analogous with respect to a matter wave subject to a static magnetic field is discussed in Sec.~\ref{sec:pseudo_spin}. The polarization structuring of the beam and the dependence on the material birefringence  is investigated in Sec.~\ref{sec:paraxial_simulations} and Sec.~\ref{sec:birefringence_role} respectively, using a paraxial model in the rotated framework where the dielectric tensor is locally diagonal. In Sec.~\ref{sec:spin_orbit} the interpretation in terms of spin-orbit interaction and inhomogeneous magnetic field in the reciprocal space is finally provided. 

\section{The basic idea and the intuitive explanation of the effect}
\label{sec:basic_idea}

Let us consider a twisted anisotropic material, where the rotation angle varies only along one direction, $x$. The optic axis is locally determined by a vectorial field called director, denoted with $\hat{n}$. In this Paper we will consider only rotation of the optic axis in the plane normal to the initial wavevector, the latter taken parallel to the axis $z$. The local rotation is determined by the angle $\theta(x)$ formed by the optic axis with $\hat{y}$, in turn providing $\hat{n}=\left(\sin\theta,\cos\theta,0 \right)$ in Cartesian coordinates. The rotation angle $\theta$ is supposed to be odd on the transverse plane, $\theta(x)=-\theta(-x)$. We start our considerations by taking a circularly-polarized plane wave impinging on the material with a wavevector parallel to the axis $z$. Polarization will then change in a different pathway in any point $x$. When the length of the material is such that the phase delay between the principal components is $\pi$ (i.e., the half-wave plate condition), beam polarization will become once again homogeneous along $x$, but with an inverted helicity with respect to the input state. Jones formalism \cite{Jones:1941}, which neglects the action of the diffraction operator, shows that also an additional phase delay $\Delta \phi_\mathrm{geo}(x)= \pm 2\theta(x)$ appears, the sign being fixed by the input helicity \cite{Bhandari:1997,Bomzon:2001_2}. Ultimately, such a counter-intuitive delay is associated with a peculiar type of geometric phase \cite{Berry:1984}, first discussed by Pancharatnam \cite{Pancharatnam:1956}. Going back to the behavior of light after crossing a structured anisotropic material, left and right circular polarizations (LCP and RCP, respectively) undergo a phase gradient opposite in sign, thus being deflected at the exit of the slab of anisotropic material along opposite directions. This idea has been largely investigated in the last few years to realize polarization gratings \cite{Gori:1999,Provenzano:2006}, in some cases called polarization-dependent prisms/beam splitters \cite{Hasman:2002} and interpreted as a SHE \cite{Ling:2015}. 

The primary question addressed in this Paper revolves around light propagation in a long anisotropic structured material, that is, extending well beyond the half-wave plate length. We consider an anisotropic material where the two eigenvalues of the dielectric tensors $\epsilon_\bot=n_\bot^2$ and $\epsilon_\|=n^2_\|$ are homogeneous, but the direction of the optic axis varies only along one direction, dubbed $x$. The material birefringence is $\Delta n = n_\|-n_{\bot}$. We consider monochromatic fields of wavelength $\lambda$ and wavenumber $k_0=2\pi/\lambda$. 
 We start by using a generalized formula for the phase difference between two electromagnetic waves $\bm E_1$ and $\bm E_2$ (or two portions of the same wave) valid also when the polarization is not identical. Following Pancharatnam's original intuition, Berry proposed the formula $\Delta \phi = \text{arg} \left( \bm E_1 \cdot \bm E_2^* \right)$, where the asterisk indicates the complex conjugate \cite{Berry:1994}.
Such formula is applied to two circularly-polarized waves propagating in two slabs made of the same anisotropic material, but rotated with respect to each other by an angle $\Delta\theta$. The relative phase delay for small $\Delta \theta$ is then 
$\Delta \phi(\delta) = \pm 2 \Delta \theta \sin^2\left( \frac{\delta}{2}\right)$,
where $\delta = k_0 \Delta n z$ is the phase retardation associated with the birefringence. By straightforward generalization to a continuously twisted material, we find
\begin{equation} \label{eq:wavefront}
    \phi(x,z) = \pm 2 \theta(x) \sin^2\left(\frac{k_0\Delta n z}{2}\right) .
\end{equation}
Eq.~\eqref{eq:wavefront} tells us that the transverse phase modulation in the wavefront is proportional to twice the rotation angle $2\theta(x)$, but it is also modulated periodically in propagation. The periodic modulation along the propagation direction $z$ is featuring a non-vanishing averaged value: we thus expect that optical beams of finite size accumulate a transverse shift while propagating in the twisted material, with a direction of the transverse motion (basically, left or right deflections) dependent on the initial helicity. The local wavevector $\bm k_l(x,z)$ is given by the gradient of the phase front; using Eq.~\eqref{eq:wavefront} in the definition of $\bm k_l(x,z)$  yields 
\begin{equation}  \label{eq:transverse_wavevector}
    \bm k_l \cdot \hat{x}= \pm 2 \sin^2\left(\frac{k_0\Delta n z}{2}\right) 
    \frac{\partial \theta}{\partial x}.
\end{equation}
In the paraxial limit, due to the fact that the walk-off stays negligible for beams propagating forming a small angle with $\hat{z}$, the wavevector is parallel to the Poynting vector. We first call $x_0(z)$ the position of the beam at each $z$: in an isotropic material encompassing a refractive index $n(x,z)$ geometrical optics (or equivalently the Ehrenfest theorem) provides $d^2x_0/dz^2=\partial_x n/{n_0}$, where 
 ${n_0}$ is the background value of $n(x,z)$. Remembering that $dx_0/dz=k_x/k_z\approx k_x/(k_0n_0)$, in the limit of invariant $\partial_x\theta$ along $z$ (see Appendix~\ref{app:trajectory} for the solution in the general case) we find
\begin{equation} \label{eq:trajectory}
    x_0(z) = \pm \frac{1}{k_0 \overline{n}}\frac{\partial \theta}{\partial x} \left[z - \frac{\sin\left(k_0 \Delta n z \right)}{k_0\Delta n} \right],
\end{equation}
where $\overline{n}=\left. \left(n_\bot + n_\| \right)\right/2$ is the refractive index perceived by circular polarizations.
\begin{figure}
    \centering
  \includegraphics[width=0.99\linewidth]{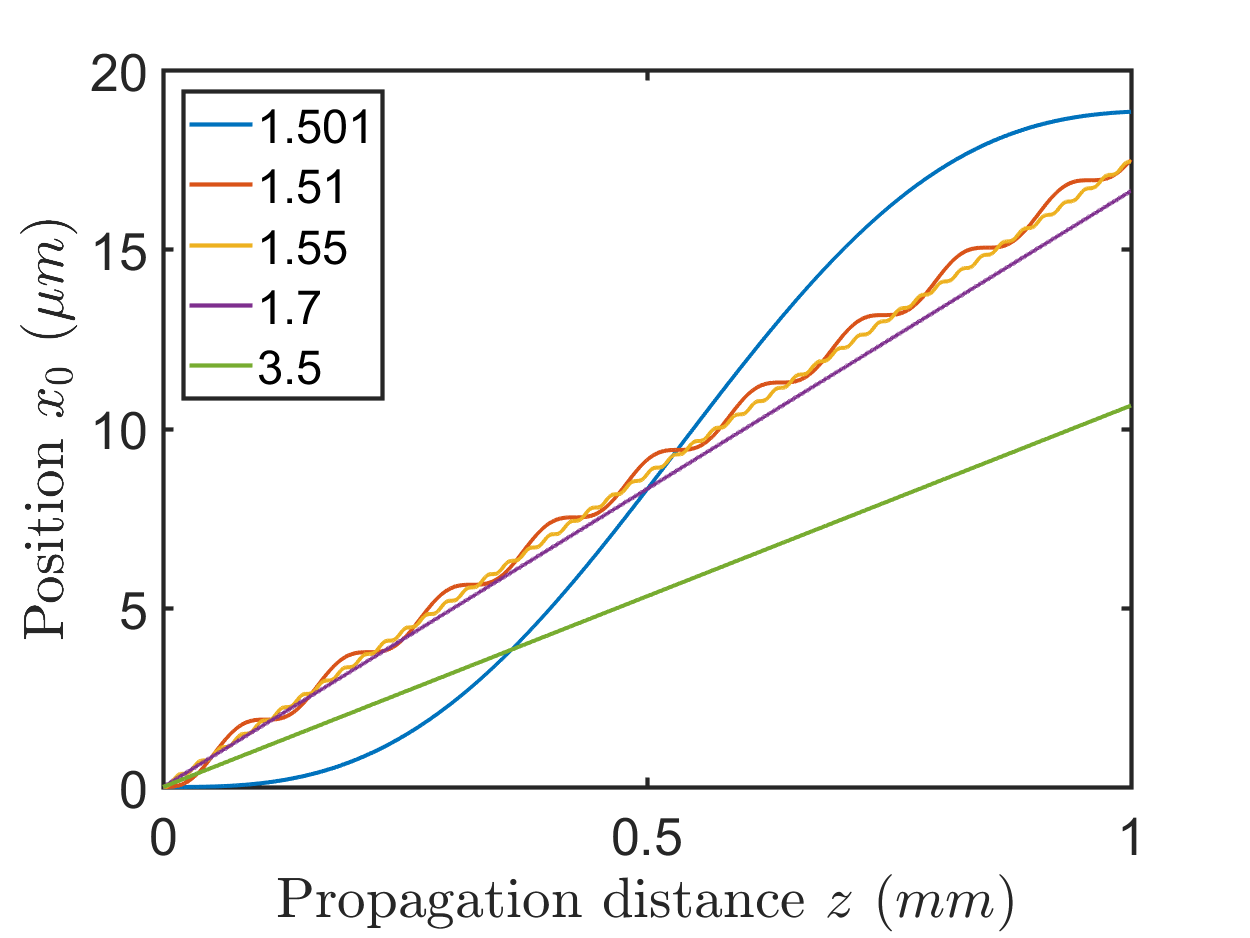}
    \caption{ Beam trajectories on the propagation plane $xz$ predicted from Eq.~\eqref{eq:trajectory}, for different values of $n_\|$ (reported in the legend) and for $n_\bot=1.5$ fixed. The twisting angle is $\theta = \theta_0 x/L$, with $\theta_0=360^\circ$ and $L=40~\mu$m; wavelength is 1064~nm. Only one helicity is plotted: trajectories for the other CP are simply found changing the sign of $x_0$. }
    \label{fig:theory_trajectories}
\end{figure}
Equation~\eqref{eq:trajectory} firmly establishes the physical phenomenon at the core of this Paper: the beams move along a tilted trajectory, with a slope dictated by the local gradient $\partial_x \theta$ in the rotation angle. Beam paths for a wavelength of $1.064~\mu$m are shown in Fig.~\ref{fig:theory_trajectories}. The beams bend towards the left or right direction according to their input helicity, with the two trajectories being mirror-symmetric to conserve the overall parity of the system. Our simple model predicts that the average deflection angle does not depend on the medium birefringence $\Delta n$, but only on the geometrical properties of the sample and on the average refractive index $\overline{n}$. The oscillating term in the trajectory decreases either if the material birefringence is increased (see Fig.~\ref{fig:theory_trajectories}) or shorter wavelengths are employed. The model implicitly assumes that the spreading due to diffraction is negligible over one birefringence length: for small birefringence we expect that Eq.~\eqref{eq:trajectory} fails to accurately capture the full wave-like propagation of the field. Despite that, in the limit $\Delta n \rightarrow 0$ the beam propagates straight: as a matter of fact, in an isotropic material the twisting cannot affect the wave evolution. Examples of the role played by diffraction can be found in Ref.~\cite{Jisha:2019} in the case of a full-wave plate long sample. 

Several proposals of PSHE are plagued by tiny polarization-dependent transverse shifts, often in the order of the wavelength itself \cite{Hosten:2008,Zhou:2019}. Regarding the spin-dependent deflection discussed here, Eq.~\eqref{eq:trajectory} allows the estimation of the expected deflection in a straightforward manner. To fix the ideas, let us focus on the case $n_\bot=1.5$ and $n_\|=1.7$ for this discussion (violet curve in Fig.~\ref{fig:theory_trajectories}). Already from Fig.~\ref{fig:theory_trajectories}, the shift is typically several times the wavelength for thick enough samples. For a more realistic length of 100~$\mu$m, the shift decreases to 2~$\mu$m. The slope of the trajectories is directly proportional to $\theta_0$ and inversely proportional to $L$. Thus, for an angle $\theta_0=45^\circ$, deflections plotted in Fig.~\ref{fig:theory_trajectories} reduce to $25\%$. On the other side, smaller $L$ greatly increases the slope: for $L=2~\mu$m a slope of $2.4^\circ$ and $19^\circ$ is achieved at $\theta_0=45^\circ$ and $\theta_0=360^\circ$, respectively. 

In our previous work \cite{Jisha:2017_1} we showed that, in the geometry investigated here, light perceives an effective potential proportional to $(\partial_x\theta)^2$ regardless of its input polarization, arising from the so-called Kapitza effect. In essence, the periodic modulation of the phase [see Eq.~\eqref{eq:transverse_wavevector}] implies a periodic modulation of the kinetic energy \cite{Kapitza:1951}, the latter having a non-vanishing average due to the quadratic relation with respect to $k_x^2$. After this initial explanation based upon ray optics, in this Article we will improve the old model based upon the Kapitza effect by describing the full polarization structure of the wave in propagation, showing how the input handedness can be used to select different optical paths. 

In our treatment we implicitly assumed that the gradient in the twist angle does not change as the beam shifts its position along the transverse direction $x$. This assumption is true only when $\theta$ is linear versus $x$, $\theta(x)=\alpha x$. For a more generic distribution of the twisting angle, higher order effects will take place, ranging from a $z-$dependent bending in the wavefront, to the emergence of more complex polarization structures \cite{Jisha:2023}. We will refine in the following sections the model by a mix of numerical simulations and interpretation in terms of fictitious magnetic fields. 

\section{Verification via FDTD numerical simulations}
\label{sec:FDTD}
Whereas thin twisted anisotropic materials have been manufactured in the last two decades using different technologies including metasurfaces \cite{Karimi:2014,Kamali:2018}, patterned liquid crystals \cite{Kim:2015,Rubano:2019} and laser-written nanogratings \cite{Drevinskas:2017}, currently it is quite challenging to manufacture a twisted anisotropic material long enough to permit the direct observation of the PSHE we described above. We thus resort to full numerical simulations of the Maxwell's equations based upon a FDTD algorithm, which can be considered as \textit{ab initio} simulations for electromagnetic waves in the classical regime. Specifically, we used the open access software MEEP \cite{Oskooi:2010}. If not stated otherwise, hereafter we will make the following ansatz for the rotation angle:
\begin{equation}
    \theta(x)= \theta_0 \tanh\left(\frac{x}{L} \right).
\end{equation}
\begin{figure}
    \centering
  \includegraphics[width=0.99\linewidth]{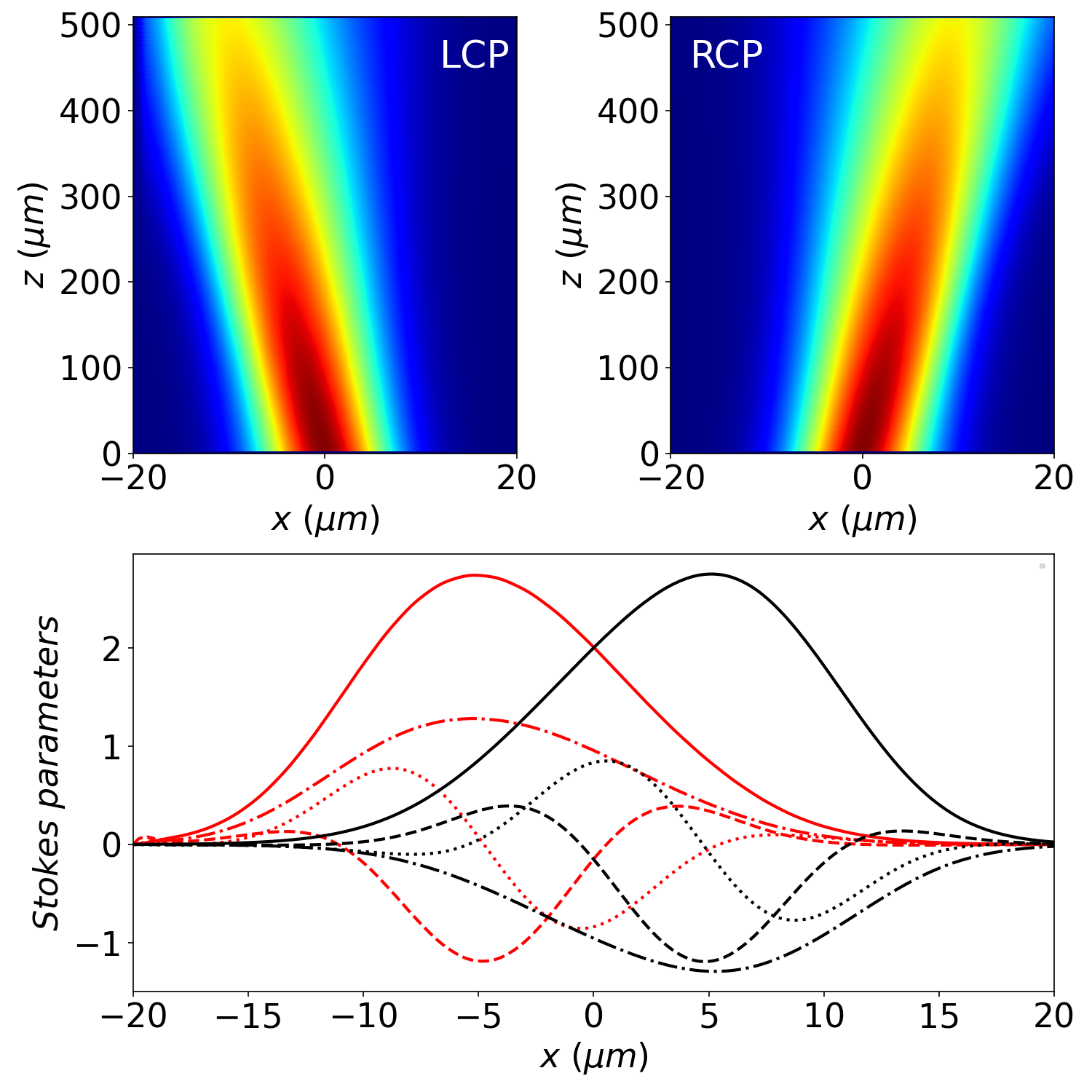}
    \caption{FDTD simulations for CP inputs when $\theta_0=180^\circ$, $L=20~\mu$m, and an input Gaussian beam of width 10~$\mu$m. Top row: distribution of the $z$-component of the Poynting vector on the plane $xz$ for LCP (left side) and RCP (right side) inputs. Bottom panel: behavior of the non-normalized Stokes parameters versus $x$ at $z=250~\mu$m for RCP (red lines) and LCP (black lines) inputs. Solid, dashed, dotted, dash-dotted curves correspond to $S_0$ (i.e., the intensity), $S_1$, $S_2$, and $S_3$, respectively. Here $n_\bot=1.5$ and $n_\|=1.7$. }
    \label{fig:figure_FDTD_mirror_symmetry}
\end{figure}
Although the details of the beam trajectories depend on the shape of $\theta$ (see below), the spin-dependent deflections is insensitive to the selected shape. Hyperbolic tangent is thus a perfect candidate to model a step-like jump in the twisting angle. \\
We start by simulating the optical propagation when a Gaussian beam of width $10~\mu$m and circular polarization is injected into the twisted material with $L=20~\mu$m and $\theta_0=180^\circ$. 
\begin{figure}
    \centering
  \includegraphics[width=0.99\linewidth]{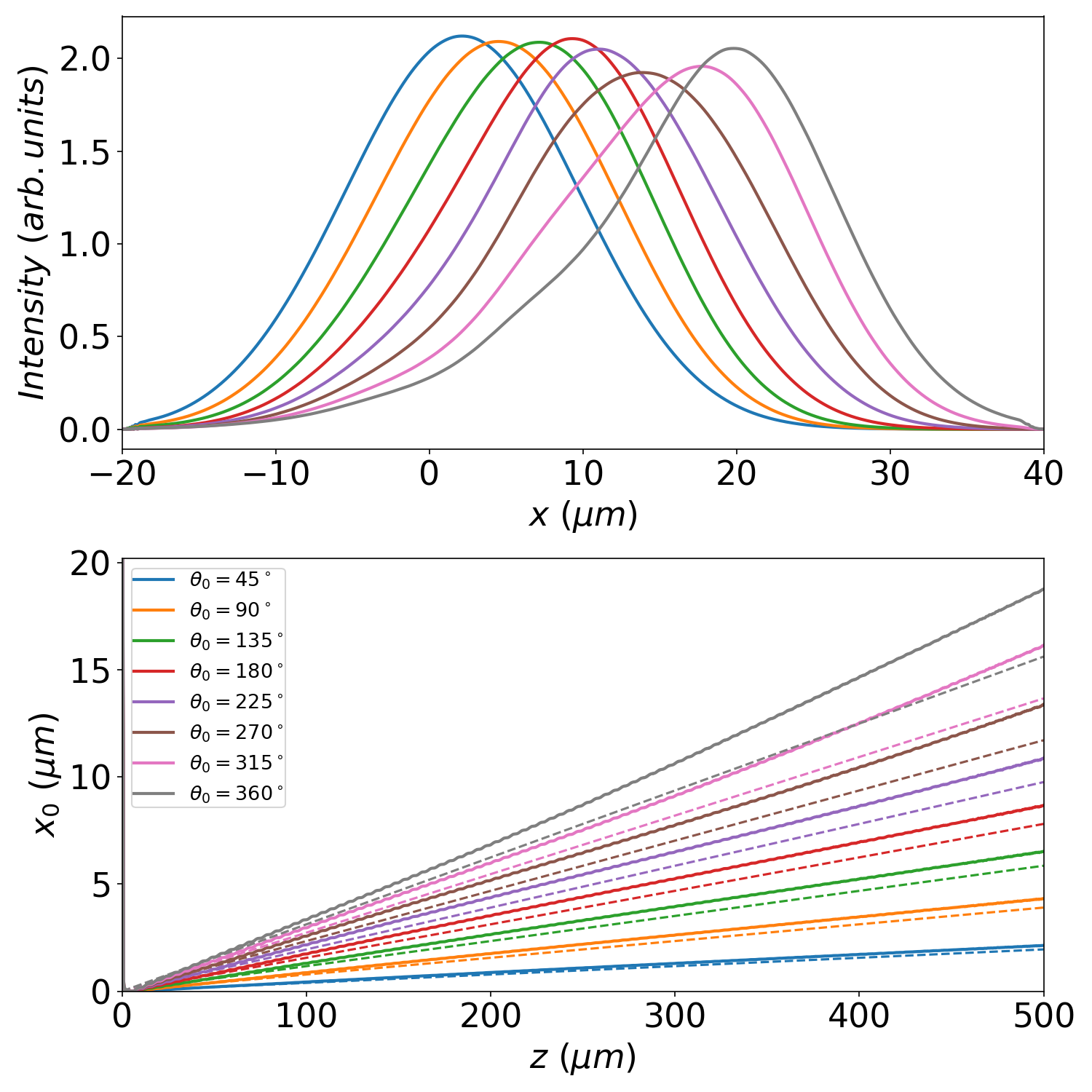}
    \caption{FDTD simulations versus the rotation angle $\theta_0$. Cross-section of the Poynting vector sampled in $z=500~\mu$m (top panel) and the corresponding trajectories on the plane $xz$ (solid lines in the bottom panel) for different values of $\theta_0$ as provided by the legend. Dashed lines on the bottom panel are the theoretical predictions from Eq.~\eqref{eq:trajectory}.  Here input is RCP, $L=20~\mu$m, $n_\bot=1.5$ and $n_\|=1.7$.}
    \label{fig:figure_FDTD_vs_theta0}
\end{figure}
Figure~\ref{fig:figure_FDTD_mirror_symmetry} shows the comparison for fields at the entrance of opposite helicity, i.e., LCP (left column) and RCP (right column). In agreement with the theory, beams are perceiving a net transverse shift, whose sign depends on the initial handedness. The average trajectory is almost rectilinear. A small periodic wiggling around the average trajectory with a period equal to $\Lambda$ is present, even if not visually perceivable in the scale of the plotted figure. Due to the birefringence of the material, the Stokes parameters oscillate periodically along $z$. In first approximation, the local values of the Stokes parameters follow the local rotation of the optic axis, presenting nodal lines (i.e., vanishing values) parallel to the axis $z$ where either $S_1$ (zones where $\theta=m\pi$) or $S_2$ (zones where $\theta=\pi/4 + m\pi$) are vanishing.  When the input handedness is switched, the Stokes parameters  transform as
\begin{align}  \label{eq:stokes_symmetry_transformation_S1}
    S^{RCP}_1(x) &=  S^{LCP}_1(-x), \\
    S^{RCP}_j(x) &= - S^{LCP}_j(-x), \ (j=2,3).
    \label{eq:stokes_symmetry_transformation_S23}
\end{align}
As a matter of fact, a switch in the handedness corresponds to a mirror reflection with respect to $x=0$ due to the odd parity of $\theta$. \\
A more quantitative comparison with the theory can be performed if we consider the behavior of the beam with $\theta_0$, that is, as the gradient in $\theta$ gets larger. The profiles of the Poynting vector in $z=0.5~$mm and the trajectories $x_0$ are plotted in Fig.~\ref{fig:figure_FDTD_vs_theta0}. As predicted, the beam shift is proportional to the maximum rotation angle $\theta_0$. At the same time, the cross-section of the beam is deformed as the gradient increases. Simultaneously with such a partial reshaping, the trajectories move away from the theoretical predictions provided by Eq.~\eqref{eq:trajectory}. The discrepancy will be explained below in detail: we can anticipate that this is due to the inhomogeneous gradient in the twisting angle seen by the beam as it propagates along $z$. \\
\begin{figure}
    \centering
  \includegraphics[width=0.99\linewidth]{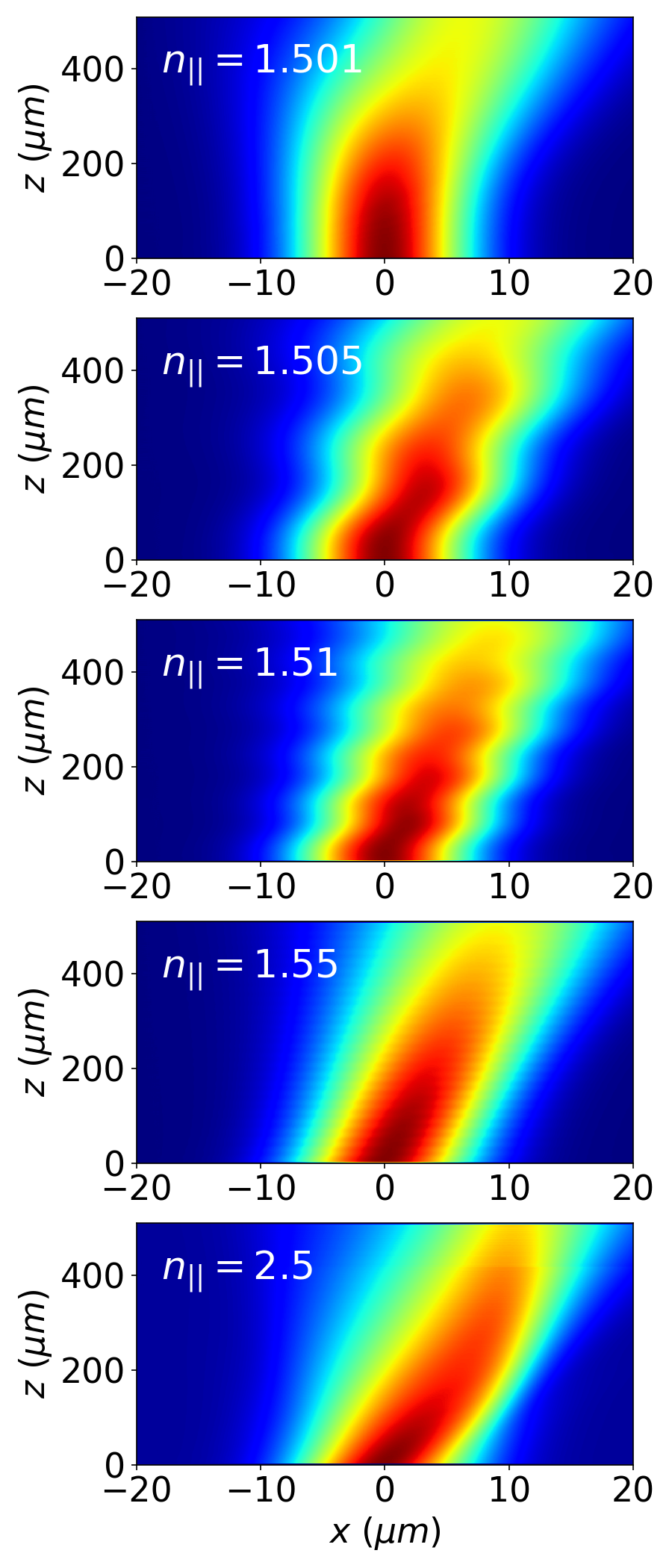}
    \caption{Distribution over the plane $xz$ of the component along $z$ of the real part of the Poynting vector versus the material anisotropy as computed by FDTD simulations. Here $n_\bot=1.5$, $L=20~\mu$m, and $\theta_0=180^\circ$. }
    \label{fig:figure_FDTD_vs_anisotropy}
\end{figure}
We finally test the behavior of a Gaussian beam as the material anisotropy $\epsilon_a$ is varied. Results are shown in Fig.~\ref{fig:figure_FDTD_vs_anisotropy}. In agreement with Fig.~\ref{fig:theory_trajectories}, the beam is transversely shifted of a very similar amount, regardless of the medium birefringence; the biggest difference is indeed a periodic wiggling of the beam, with a period corresponding to the birefringence length in agreement with Eq.~\eqref{eq:trajectory}. Such a periodic transverse motion significantly impacts on the intensity profile only when the birefringence length is comparable with the Rayleigh length of the beam, that is, a snake-like behavior takes places at low anisotropies.

\section{Pseudo-spin in the presence of a Zeeman-like interaction}
\label{sec:pseudo_spin}
Owing to the definition of $\theta$, the dielectric tensor elements are then $\epsilon_{ij}(x)=\delta_{ij}\epsilon_\bot+\epsilon_a n_i(x) n_j(x)\ (i,j=x,y,z)$, where $\epsilon_a=\epsilon_\|-\epsilon_\bot$ is the optical anisotropy, $n_i$ are the components of the director field, and finally $\delta_{ij}$ is the Kronecker delta \cite{Simoni:1997}. As discussed above, PSHE requires an odd symmetric distribution for the rotation angle, that is, $\theta(x)=-\theta(-x)$. 
In previous works \cite{Jisha:2017_1} it has been theoretically demonstrated that even distributions can yield to optical waveguides based upon the geometric phase, but polarization-independent, due to the Kapitza effect. \\
The finite-size optical beams of interest are supposed wide enough to be in the paraxial limit. The electromagnetic field can then be described as a two-component vector $\bm \psi=\left(E_x;\ E_y \right)$. The Maxwell equations reduce  to \cite{Jisha:2023}
\begin{equation}  \label{eq:spinorial_E}
 \nabla^2 \bm \psi + k_0^2 \left\{\overline{\epsilon} \bm I + \frac{\epsilon_a}{2} \left[ \bm \sigma_1 \sin(2\theta) - \bm \sigma_3 \cos(2\theta)\right] \right\}\cdot \bm \psi = 0,   
\end{equation}
where $\bm \sigma_i\ (i=1,2,3)$ are the Pauli matrices and $\overline{\epsilon}= \left(\epsilon_\bot + \epsilon_\| \right)/2$. Equation~\eqref{eq:spinorial_E} shows the close similarity between the optical propagation in a twisted anisotropic medium and the propagation of charged particles in a magnetic field. 
Introducing the standard vector of Pauli matrices $\bm \sigma$, the Hamiltonian term accounting for the magnetic interaction can be indeed written in the form 
\begin{equation}
  H_{Zeeman}= \frac{\epsilon_a}{2} \left[ \bm \sigma_1 \sin(2\theta) - \bm \sigma_3 \cos(2\theta)\right]   = \frac{1}{2}   \bm{\sigma} \cdot \bm B_\mathrm{eff}, \label{eq:magnetic_field}
\end{equation}
where $\bm B_\mathrm{eff}=\epsilon_a\left[ \sin(2\theta) \hat{e}_1 - \cos(2\theta) \hat{e}_3\right]$ represents an effective magnetic field \cite{Kuratsuj:1998,Fang:2013,Rechtsman:2013_1,Schine:2016}, here defined within a three-dimensional vector space spanned by unit vectors $\hat{e}_i\ (i=1,2,3)$. In doing that, we assumed the equivalent magnetic moment to be unitary. For vanishing $\epsilon_a$, Eq.~\eqref{eq:magnetic_field} turns into the well known Helmholtz equation for light propagating in an isotropic material. In close analogy with the Zeeman effect, a non-vanishing anisotropy $\epsilon_a$ breaks the degeneracy between the ordinary and the extraordinary component, now propagating with a different refractive index. With respect to the standard SG experiment, the gradient in the Zeeman-like potential energy stems from a spatially-varying rotation of the pseudp-magnetic field, whereas its magnitude is uniform along $x$.  

According to Eq.~\eqref{eq:magnetic_field}, the spin flips with a period $\Lambda$ fixed by the birefringence $\Delta n=n_\| - n_\bot $, $\Lambda=\lambda/\Delta n$, thus remaining constant along the transverse direction $x$. The equivalent force acting on the wave is periodic along $z$, yielding a vanishing net-shift along the transverse plane. On the other side, a spin-dependent odd phase modulation proportional to $2\theta$ appears at $z=\Lambda/2$, which has been extensively used to demonstrate spin-dependent deflectors or polarization gratings once these beams propagate out of the structured material \cite{Gori:1999,Hasman:2002,Provenzano:2006,Ling:2015}. Such a behavior is therefore in agreement with our intuitive model discussed in Sec.~\ref{sec:basic_idea}.

Both the intuitive explanation (see Sec.~\ref{sec:basic_idea}) and the full numerical simulations shown in Sec.~\ref{sec:FDTD} show a prominent role played by the two circular polarizations. We thus try to rewrite Eq.~\eqref{eq:spinorial_E} using the LCP and RCP as basis for expressing the field $\bm \psi$. Calling $\bm{\psi}_{LR}$ the field in the CP basis (see Appendix~\ref{app:transformation_to_circular} for details about the transformation), we find that 
\begin{multline}  \label{eq:Maxwell_no_longitudinal_LR} 
 {\nabla^2 \bm \psi_{LR} } +  k_0^2\overline{\epsilon}  \bm \psi_{LR} + \\
  k_0^2 \epsilon_a \left[ \bm \sigma_2 \sin(2\theta) - \bm \sigma_1 \cos(2\theta)\right] \cdot \bm \psi_{LR} = 0. \end{multline}
Equation~\eqref{eq:Maxwell_no_longitudinal_LR} is of the same form of Eq.~\eqref{eq:spinorial_E}. The most important difference is that the effective magnetic field is now directed along $e_1$ and $e_2$, a clear major difference with respect to the case of matter waves subject to a magnetic field.

\section{Paraxial model and simulations in the rotated framework}
\label{sec:paraxial_simulations}

It thus seems that  we are in the presence of a magnetic field inhomogeneity only in the real space, furthermore providing a vanishing transverse momentum to the photons when averaged over an integer number of birefringence periods $\Lambda$. An alternative and more complete view on the system can be achieved when Eq.~\eqref{eq:magnetic_field} is rewritten in a reference system where the dielectric tensor is diagonal in each point. The required point-wise rotation is
\begin{equation}  \label{eq:rotation}
\bm \psi^\prime = \bm{R}(\theta) \cdot \bm \psi = \bm e^{i\bm \sigma_2 \theta(x)} \cdot \bm \psi. 
\end{equation}
A similar kind of rotation has been implemented in Ref.~\cite{Sukumar:1997} to address the spin-flip of atoms confined by magnetic traps.
Furthermore, such an approach is conceptually similar to Ref.~\cite{Bliokh:2008_1}, where the case of a beam propagating along 3D trajectories in an isotropic material was investigated: a coordinate transformation such to align everywhere the framework to the local wavevector was employed to find the expression for the spin-orbit interaction, in that case stemming from a Coriolis-like force. The major difference is that here we are using a transformation acting on the field itself, whereas the spatial reference system stays untouched. 

Starting from Eq.~\eqref{eq:spinorial_E}, we find the vectorial paraxial equation \cite{Slussarenko:2016,Jisha:2017_1,Jisha:2017}
\begin{multline}
  2ik_0 \bm N \cdot \frac{\partial \bm{u}}{\partial z}=   \\
 -   \frac{\partial^2 \bm{u}}{\partial x^2} + \left(\frac{\partial\theta}{\partial x} \right)^2  \bm{u}   + i \frac{\partial^2 \theta}{\partial x^2}  \bm{\tilde{\sigma}}_2\cdot \bm{u} + 2i  \frac{\partial\theta}{\partial x} \bm{\tilde{\sigma}}_2 \cdot \frac{\partial \bm{u}}{\partial  x},  \label{eq:maxwell_rotated_inho_paraxial}
\end{multline}
where $\bm u$ is the slowly varying envelope, whose components are parallel in each point to the local principal axes of the crystal thanks to the rotation determined by Eq.~\eqref{eq:rotation}. We further defined the diagonal matrix $\bm N = \left(n_\bot,0;0,n_\| \right)$ and the modified Pauli matrix $\bm{\tilde{\sigma}}_2 (z)= e^{-ik_0\bm N z} \cdot \bm \sigma_2 \cdot e^{ik_0\bm N z}$. Even if written in a rotated system such to follow the local axes of the material, Eq.~\eqref{eq:maxwell_rotated_inho_paraxial} includes $z$-dependent terms through $\bm{\tilde{\sigma}}_2$, essentially due to the coupling between different transverse positions intrinsic to diffraction.\\
\begin{figure}
    \centering
  \includegraphics[width=0.99\linewidth]{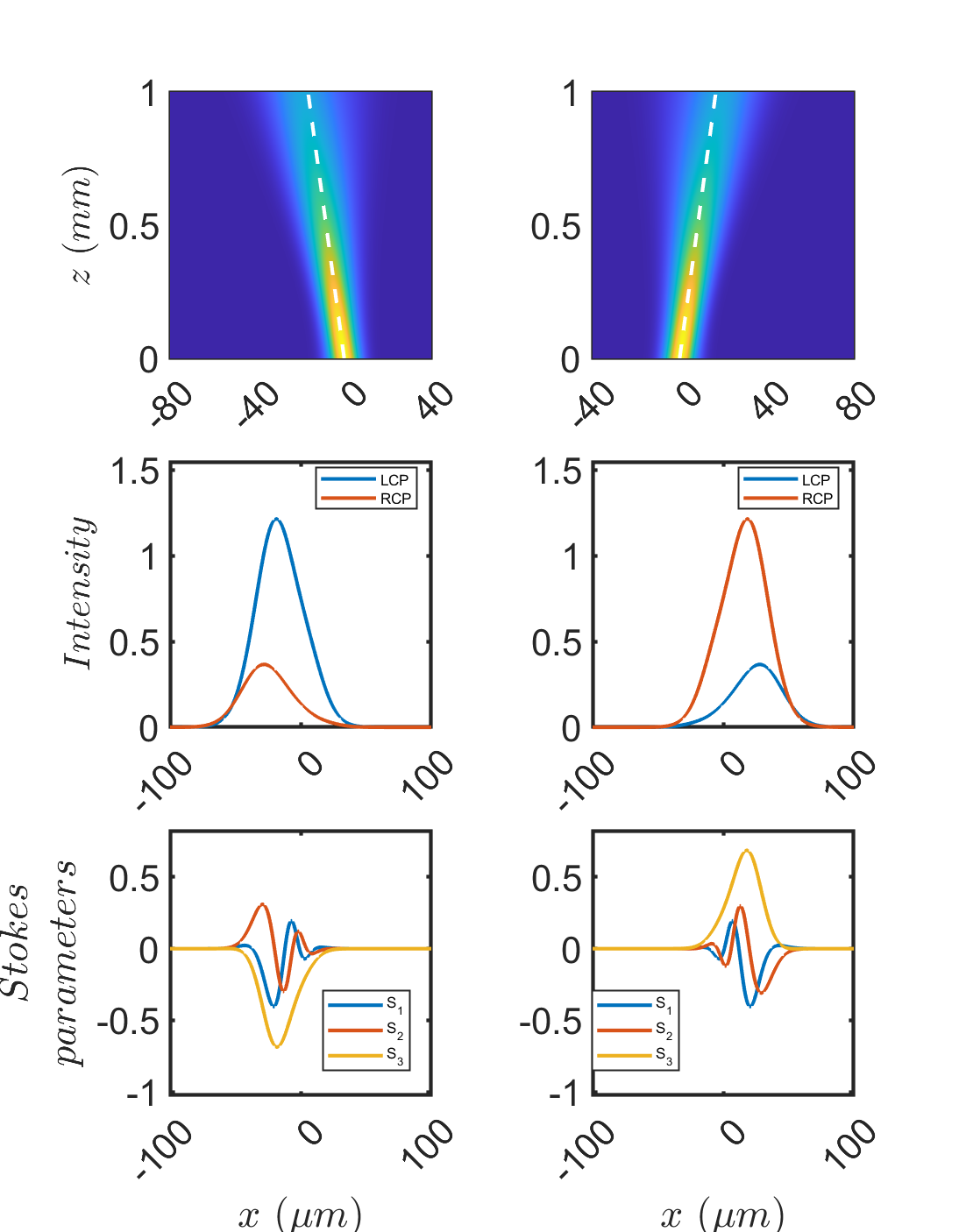}
    \caption{ Propagation calculated using Eq.~\eqref{eq:maxwell_rotated_inho_paraxial} for a Gaussian beam  LCP (left column) and RCP (right column), an input waist $10~\mu$m, and $\lambda=1064~$nm. First row: intensity $|u_x|^2+|u_y|^2$ on the plane $xz$; the white dashed lines are the theoretical trajectories according to Eq.~\eqref{eq:trajectory}. Second (third) row: intensity of the LCP and RCP (non-normalized Stokes components) components calculated versus $x$ at $z=1~$mm.  The material parameters are $n_\bot=1.5$, $n_\|=1.7$, $\theta_0=180^\circ$, and $L=20~\mu$m.}
    \label{fig:BPM_symmetry}
\end{figure}
We simulated the vectorial paraxial equation~ \eqref{eq:maxwell_rotated_inho_paraxial} using a BPM code described in Appendix~\ref{app:derivation_paraxial_model}. We start by discussing a generic example showing the mirror symmetry of the optical propagation when the input helicity is inverted, the results being shown in Fig.~\ref{fig:BPM_symmetry}. PSHE is evident, with the beam moving along a rectilinear trajectory with a slope of around $1^\circ$, yielding a transverse shift of around $18.5~\mu$m at $z=1~$mm. When helicity is switched, the electromagnetic field undergoes an exact mirror reflection with respect to the propagation axis $z$. With respect to the simple model elaborated in Sec.~\ref{sec:basic_idea}, the CP components are non-vanishing inside the material, despite a single helicity is launched into the sample both. We thus define the intensity on the two CPs as $I_{RCP}=|\bm \psi_{LR}(1)|^2$ and $I_{LCP}=|\bm \psi_{LR}(2)|^2$, plotted in the middle row of Fig.~\ref{fig:BPM_symmetry}: their transverse distribution is different, thus demonstrating an additional structuring of the beam along the transverse plane. In particular, the Stokes parameters $S_2$ and $S_3$ switch their sign when input helicity is inverted, in agreement with  Eq.~\eqref{eq:stokes_symmetry_transformation_S23} and Eq.~\eqref{eq:Maxwell_no_longitudinal_LR}.   \\
\begin{figure}
    \centering
  \includegraphics[width=0.99\linewidth]{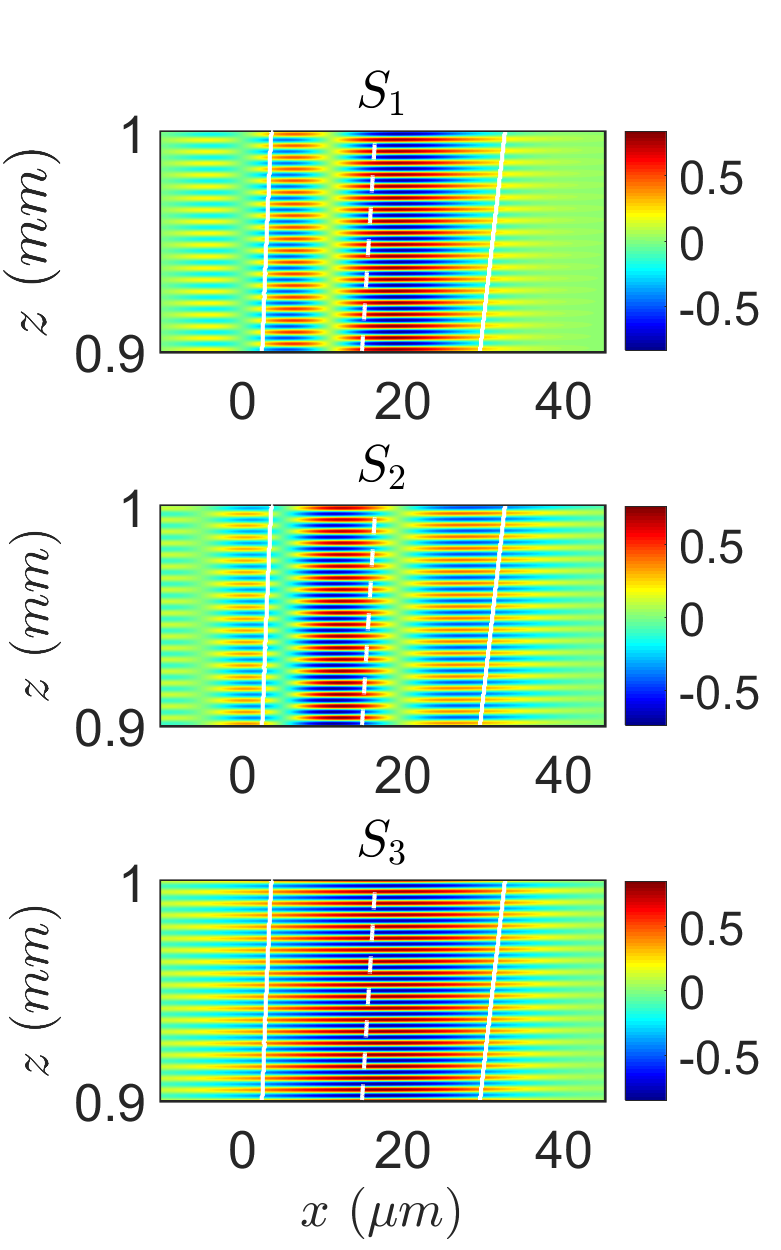}
    \caption{ Non-normalized Stokes parameters on the plane $xz$ in the interval $0.9~$mm$<z<1~$mm.  Geometry and material values are the same as in Fig.~\ref{fig:BPM_symmetry}. Input polarization is RCP. Dashed and solid white lines are the center and the FWHM positions, respectively.}
    \label{fig:stokes_xz}
\end{figure}
In agreement with the FDTD simulations (see section~\ref{sec:FDTD}), Fig.~\ref{fig:BPM_symmetry} shows a complex structured polarization, not accounted for in the simple model presented in Sec.~\ref{sec:basic_idea}. Figure~\ref{fig:stokes_xz} provides a more complete overview on the polarization, including its transverse structuring and its evolution along $z$. The Stokes parameters are oscillating periodically with the birefringence length $\Lambda/\Delta n$. Although not clearly perceived in the figure, the equi-polarization planes (i.e., the longitudinal \textit{fringes} in the plots) are slightly tilted due to the bending in the wavevector. In an anisotropic material, a CP will become linear after a quarter-wave plate distance given by $\lambda/\left( 4\Delta n\right)$, with the direction depending on the direction of the optic axis $\theta$. Accordingly, $S_1$ and $S_2$ are vanishing in the transverse positions where $\theta=m\pi$ and $\theta=\pi/4+m\pi$ ($m\in \mathbb{N}$), respectively. Summarizing, the optical propagation in the rotated system used by Eq.~\eqref{eq:maxwell_rotated_inho_paraxial} is very similar to the propagation in a homogeneous material along $z$, but the rotation $\bm R$ given by Eq.~\eqref{eq:rotation} induces periodic changes in the polarization, the inhomogeneous rotation thus being the main responsible for the transverse structuring of the beam polarization. 

\begin{figure}
    \centering
  \includegraphics[width=0.99\linewidth]{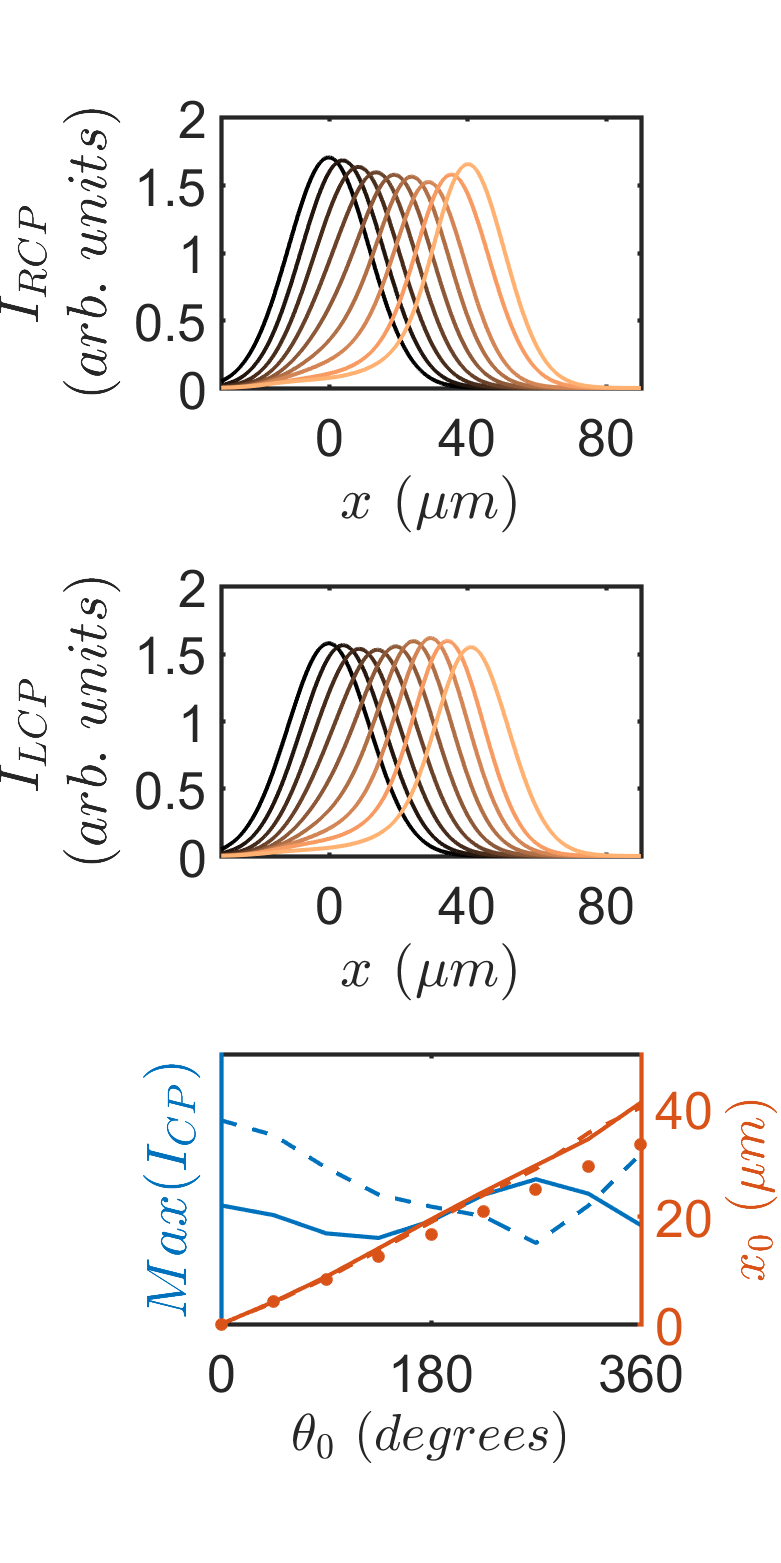}
    \caption{Behavior with respect to the maximum rotation angle $\theta_0$ at $z=1~$mm. Intensity distribution for the RCP (top panel) and LCP (middle panel) components versus $x$; each curve corresponds to $\theta_0$ from $0^\circ$ to $360^\circ$ in steps of $45^\circ$, from dark to bright respectively. (Bottom panel, left axis) Peak of the RCP (blue dashed line) and LCP (blue solid line) versus $\theta_0$. Right axis: same but the position of the components, now in red color. The red circles represent the theoretical prediction from Eq.~\eqref{eq:trajectory}. Input polarization is RCP, input beam waist is $10~\mu$m, and $L=20~\mu$m.}
    \label{fig:CP_vs_theta0}
\end{figure}

We now proceed to investigate the influence of the transverse gradient on the beam propagation. To accomplish this task, Fig.~\ref{fig:CP_vs_theta0} shows the intensity of the two CPs versus $x$ taken at their local maxima nearest to the end of the simulation window, placed at $z=1~$mm. Such cross-sections are moreover plotted for different values of the  maximum rotation angle $\theta_0$. As a matter of fact, the whole beam is rotated due to the transverse deflection (see Fig.~\ref{fig:S3_xz}), thus the longitudinal $z$ position corresponding to the cross-sections changes with the angle $\theta_0$. The shift along $x$ due to the PSHE increases with $\theta_0$, with no substantial change of energy between the two circular polarizations. A long tail on the side opposite to the deflection (i.e., negative $x$ in the figure) emerges as $\theta_0$ gets larger. Although the two CP components undergo a tiny oscillatory motion dependent on their helicity, the maxima (i.e., when $S_3$ reaches its maximum value in absolute value) are placed at the same $x-$position. The overall trend is summarized in the bottom panel of Fig.~\ref{fig:CP_vs_theta0}: for large angles $\theta_0$ the beam undergoes stronger deviations than predicted by the ray-like model providing Eq.~\eqref{eq:trajectory}.
The question now is the origin behind the transverse structuring observed in the simulations, and why the agreement in the trajectory between wave equation and the ray-theory developed in Sec.~\ref{sec:basic_idea} worsens as the $\theta_0$ gets larger, see the bottom row of Fig.~\ref{fig:CP_vs_theta0}.

\section{Dependence on the birefringence}
\label{sec:birefringence_role}
\begin{figure*}
    \centering
  \includegraphics[width=0.99\linewidth]{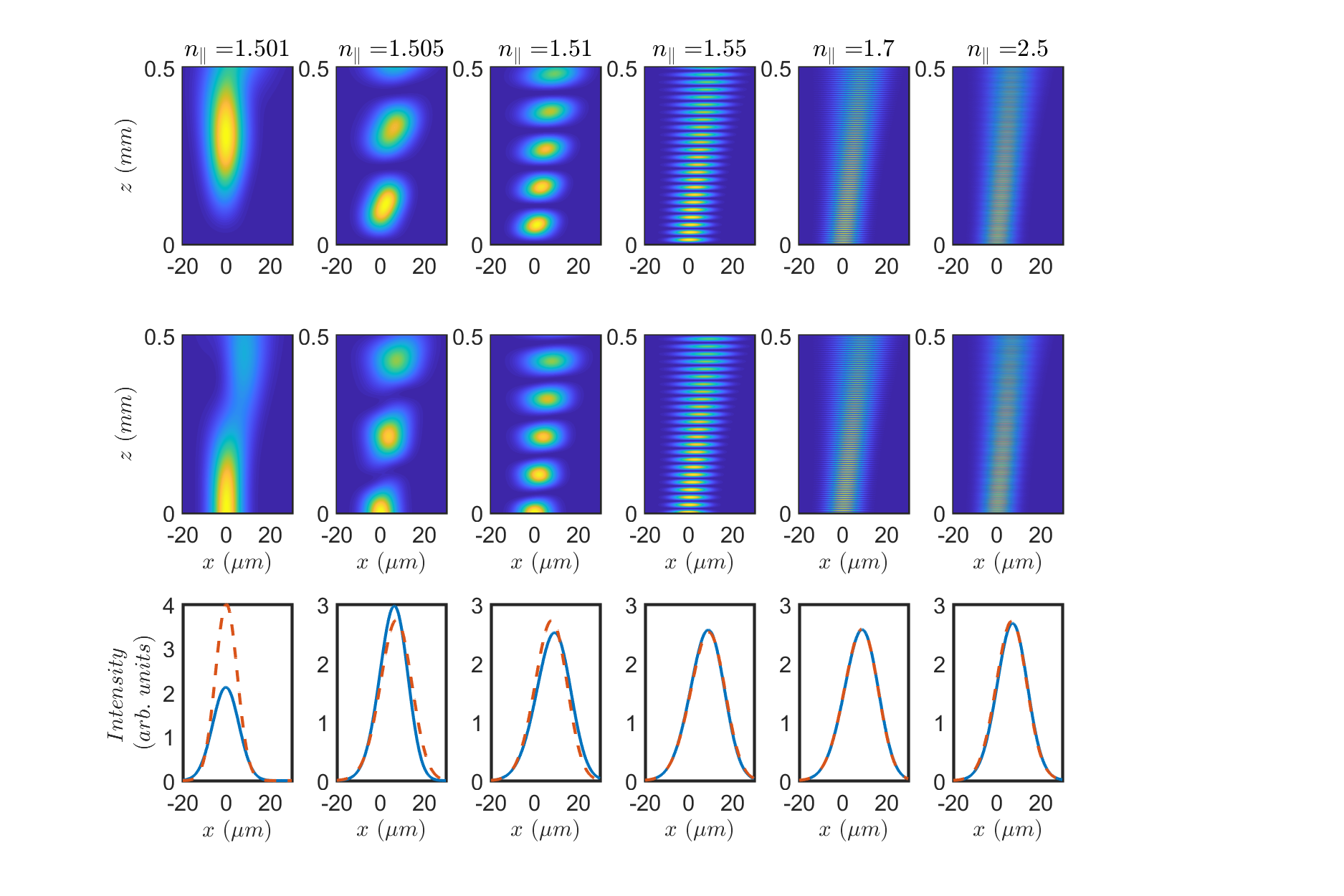}
    \caption{ Intensity distribution in the plane $xz$ for the LCP (top row) and the RCP component (middle row) when a RCP is launched at the input. Bottom row: cross-section versus $x$ cut at the closest local maximum near the output (that is, $z=0.5~$mm), for the LCP (blue solid line) and RCP (red dashed line) components. Each column corresponds to a different value for $n_\|$, shown as a title at the top. Input wavelength is 1064~nm, $w_\mathrm{in}=10~\mu$m, $n_\bot=1.5$, $\theta_0=180^\circ$, and $L=20~\mu$m. }
    \label{fig:BPM_vs_deltan}
\end{figure*}

This section is aimed to the investigation of how the wave evolution depends on the material anisotropy, that is, on the amplitude of the effective magnetic field in Eq.~\eqref{eq:magnetic_field}. The corresponding FDTD simulations have already been shown in Fig.~\ref{fig:figure_FDTD_vs_anisotropy}. Numerical simulations of the light behavior computed via Eq.~\eqref{eq:maxwell_rotated_inho_paraxial} are plotted in Fig.~\ref{fig:BPM_vs_deltan}, where the intensity carried by each CP is reported in the lab framework. In our numerical investigation we fixed $n_\bot$ and varied $n_\|$. A spin-dependent lateral deflection is observed for any value of the birefringence, accompanied by a periodic modulation of the CP components following the birefringence length. When the birefringence is too low ($\Delta n=0.001$), the beam strongly diffracts in one birefringence length, thus encompassing an appreciable transverse motion already in that propagation range. 
When $\Delta n=0.005$ and $\Delta n=0.01$, a clear PSHE is visible, but the intensity profiles still undergo appreciable transverse modifications, as clearly visible in the cross-sections. At $\Delta n=0.05$ and for $\Delta n$ up to 0.6, the two components possess a nearly identical cross-section, behaving very closely to the theoretical model discussed in Sec.~\ref{sec:basic_idea}. For further increases in $\Delta n$ (see the case $n_\|=2.5$ in Fig.~\ref{fig:BPM_vs_deltan}), the two component distribution starts to differ slightly: the larger the anisotropy the larger the difference is.
The corresponding behavior for the field $\bm u$ is presented in Appendix~\ref{app:vs_birefringence}. 

To explain the three different regimes reported in Fig.~\ref{fig:BPM_vs_deltan}, we bring Eq.~\eqref{eq:maxwell_rotated_inho_paraxial} to a more standard form. The term containing the derivative along $z$ in Eq.~\eqref{eq:maxwell_rotated_inho_paraxial} is a matrix, differently from what happens in the Pauli equation where the temporal derivative is scalar. To recast Eq.~\eqref{eq:maxwell_rotated_inho_paraxial}, we can develop $\bm N$ as a power series with respect to the normalized birefringence $\gamma=\Delta n /\overline{n}$. Following the same steps of Ref.~\cite{Jisha:2023}, up to the quadratic terms in $\gamma$ Eq.~\eqref{eq:maxwell_rotated_inho_paraxial} provides   
 \begin{multline}
        2ik_0 \overline{n} \frac{\partial \bm u}{\partial z} = \left( \bm I + \frac{\gamma}{2}\bm \sigma_3 \right) \cdot \hat{Q}(x) \bm u + 
       \bm{\tilde{\sigma}}_2 \cdot \hat{P}(x) \bm u + \\
         \frac{i\gamma}{2}\left[ \bm \sigma_2 \sin(k_0\Delta n z) - \bm \sigma_1 \cos(k_0\Delta n z) \right] \cdot \hat{P}(x) \bm u, \label{eq:propagation_SVEA_pauli_form}
\end{multline}
where we defined the operators $\hat{Q}(x)=-\frac{\partial^2 }{\partial x^2} + \left(\frac{\partial \theta}{\partial x} \right)^2$ and $\hat{P}(x) = i\left(\frac{\partial^2 \theta}{\partial x^2} + 2\frac{\partial \theta}{\partial x} \frac{\partial }{\partial x} \right)$. Equation~\eqref{eq:propagation_SVEA_pauli_form} shows that the dependence on $\Delta n$ is not solely due to the modified Pauli matrix $\bm{\tilde{\sigma}}_2$.
A direct analysis of Eq.~\eqref{eq:propagation_SVEA_pauli_form} permits us to understand the existence of the three regimes described above. For small $\gamma$, the modified Pauli matrix $\bm{\tilde{\sigma}}_2$ varies slowly along $z$  with respect to the Rayleigh distance, thus inducing a wave transverse motion already inside a birefringence length $\lambda/\Delta n$. For intermediate $\gamma$, $\bm{\tilde{\sigma}}_2$ varies fast enough to not impact significantly on the wave propagation. Once the smallness of $\gamma$ is accounted for, the optical propagation is thus mainly determined by the operator $\bm I \hat{Q}(x)$. When $\gamma$ is comparable to unity, the term proportional to the Pauli matrix $\bm \sigma_3$ turns out to be relevant, leading to a different behavior for the extraordinary and the ordinary components. More details are provided in Appendix~\ref{app:vs_birefringence}.


\section{A wave model for the spin-orbit coupling}
\label{sec:spin_orbit}
Given that the momentum operator $\hat{p}_x$ and the rotation operator $\bm{R}(\theta) = e^{i\bm \sigma_2 \theta(x)}$ do not commute, Eq.~\eqref{eq:maxwell_rotated_inho_paraxial} contains terms where the transverse gradient of the wavefunction is coupled with the spin via a Pauli matrix. Thus, in the rotated reference system an effective magnetic field inhomogeneous in the reciprocal space arises. \\
In a more formal way, Eq.~\eqref{eq:spinorial_E} after the local rotation
can then be recast in a form similar to the  Pauli equation \cite{Karimi:2012,Lin:2014_1,Abbaszadeh:2021}
\begin{multline}
    \frac{\partial^2 \bm{\psi^\prime}}{\partial z^2}  + k_0^2 \left({\epsilon}_\bot \bm I - \frac{\epsilon_a}{2} \bm \sigma_3 \right) \cdot \bm{\psi^\prime}  =  \\ \left[\bm \sigma \cdot \left( \bm p - \bm{A} \right) \right]^2 \bm{\psi^\prime} + \phi \bm{\psi^\prime} 
   , \label{eq:gauge_field}
\end{multline}
where we defined the potential vector $\bm A = i\partial_x \theta\ \hat{z}$, the scalar potential $\phi = 2\left(\partial_x \theta \right)^2$, and the momentum operator $\bm p = -i\hat{x}\partial_x$ in the 1-dimensional approximation. The RHS in Eq.~\eqref{eq:gauge_field} represent the spin-orbit interaction in our system. Thanks to the gauge transformation, the effective magnetic field in the Zeeman term is now uniform in space. \\
A clearer picture can be extracted if the same operation is carried out in the circular basis: now the rotation reads $\bm{R}_{LR}(\theta)=e^{i\bm \sigma_3 \theta}$, and Eq.~\eqref{eq:Maxwell_no_longitudinal_LR} turns into
\begin{multline}
    \frac{\partial^2 \bm{\psi^\prime}_{LR}}{\partial z^2}  + k_0^2 \left(\overline{\epsilon} \bm I + \epsilon_a \bm \sigma_1 \right) \cdot \bm{\psi^\prime}_{LR}  =  \\ \left[\bm \sigma \cdot \left( \bm p - \bm{A} \right) \right]^2 \bm{\psi^\prime}_{LR} + \phi \bm{\psi^\prime}_{LR} 
   . \label{eq:gauge_field_LR}
\end{multline}
\begin{figure}
    \centering
  \includegraphics[width=0.99\linewidth]{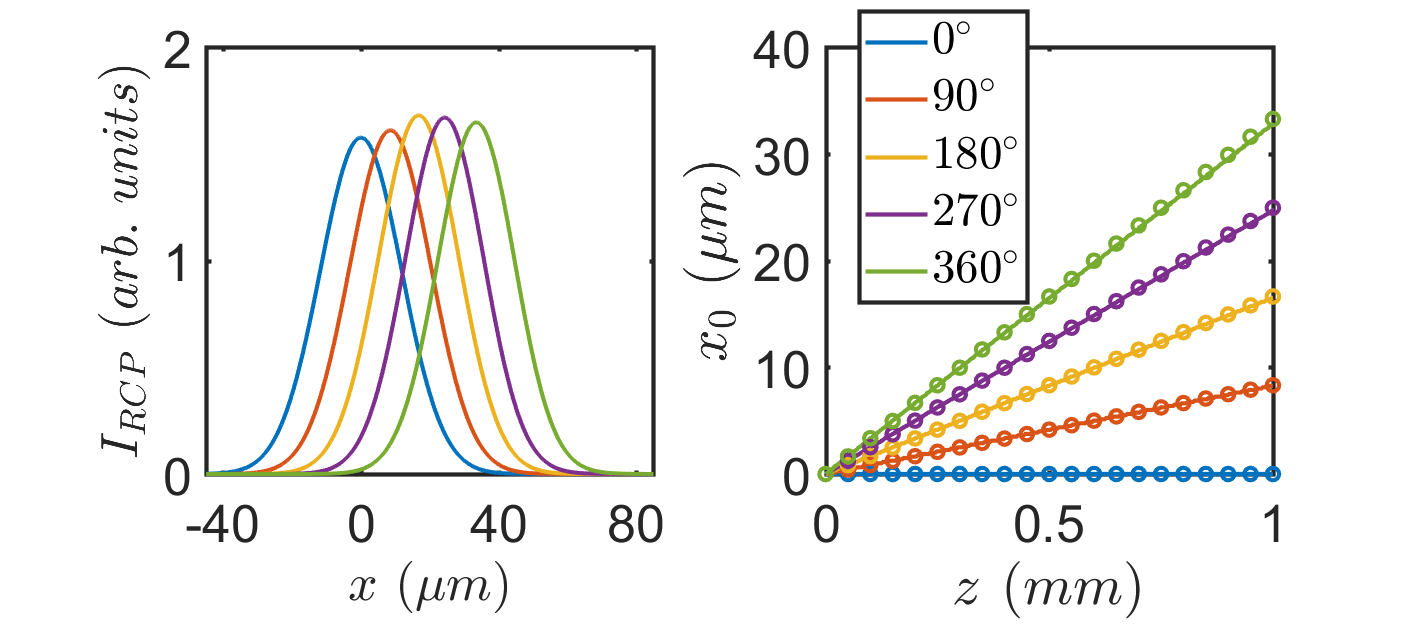}
    \caption{ Optical propagation in a linear gradient $\theta(x)=\theta_0 x/L$. Left side: Intensity profile versus $x$ of the RCP component at the maximum nearest to $z=1~$mm. Right side: beam trajectories $x_0$ versus the propagation distance $z$; circles correspond to the theoretical prediction derived from Eq.~\eqref{eq:trajectory}. Different colors correspond to different angles $\theta_0$ reported in the legend and valid for both panels. Here $\lambda=1064~$nm, $L=20~\mu$m, and $w_\mathrm{in}=10~\mu$m.}
    \label{fig:linear_gradient}
\end{figure}
The effective scalar potential and potential vector $\phi$ and $\bm A$ are unvaried, whereas the dielectric tensor differs due to the different polarization used to write down the pseudo-spinorial field. Thanks to Eq.~\eqref{eq:gauge_field} and \eqref{eq:gauge_field_LR}, we are now able to connect the intuitive explanation provided in Sec.~\ref{sec:basic_idea} to the numerical results we presented until now. The eigenvalues of the rotation operator are the two circular polarizations: when a CP is entering into the material, it will acquire a phase gradient equal to $\pm \theta$ in the rotated framework. This term provides the average slope in Fig.~\ref{fig:theory_trajectories} and in all the simulations shown in this Paper. While propagating inside the material, the potential vector $\bm A$ will act on the wave, resulting in changes in the transverse momentum $k_x$ following
\begin{equation}  \label{eq:shift_kx_gauge}
    k_x \rightarrow k_x \mp \frac{\partial \theta}{\partial x}.
\end{equation}
The Fourier transform of the field thus undergoes a simple linear phase modulation if $\theta \propto x$, resulting in a position shift in the real domain. 
The variations in the momentum in turn yield a change in the trajectory: whereas for $\theta$ linear in $x$ this is a simple change in trajectory in agreement with Sec.~\ref{sec:basic_idea}, more complicated transverse structuring for $\theta(x)$ imply all the higher-order effects observed with the numerical simulations.
Given that the polarization is continuously flipping due to the medium anisotropy, such changes in the wavevector will be periodic with period $\lambda/\Delta n$. Mathematically, the flipping is due to interaction with the terms proportional to the dielectric tensor $\bm \epsilon$ in Eq.~\eqref{eq:gauge_field_LR}, which indeed are independent from $x$.
For $\epsilon_a \rightarrow 0$, the oscillation period diverges towards infinity: the tilted wavefront stemming from Eq.~\eqref{eq:shift_kx_gauge} is perfectly compensating the tilt induced by the rotation operator at the entrance of the material, with the optical beam propagating straight, as it should be in an isotropic material.
Hence, we can safely associate $\bm A$ with the periodic oscillations along the average trajectory. 

To unveil the effects related with the wave-like nature of the field, we first consider a linear profile for $\theta$, thus implying a constant $\bm A$ and a constant scalar potential $\phi$. Typical transverse profiles of $I_{LCP}$ are plotted in Fig.~\ref{fig:linear_gradient}. Differently to the case of a non-homogeneous $\bm A$ (see Fig.~\ref{fig:CP_vs_theta0}), the beam stays Gaussian and conserves the initial parity symmetry. Nonetheless, on the plane $xz$ the Stokes parameter $S_3$ presents a tilted distribution (with respect to the $z$ axis) of its negative and positive regions related to the bending of the local wavevector, very similar to what is shown in Fig.~\ref{fig:S3_xz} in the Appendix. Similarly, the trajectory is now in perfect agreement with the theoretical prediction given by Eq.~\eqref{eq:trajectory}. This proves that a point-dependent effective potential vector $\bm A$ deforms the shape of the beam in propagation: the symmetry with respect to the beam center is broken by the variations in the phase front. Therefore, such variations are finally driven by the modulation of the local wavevector $\bm k_l(x)$, see Eq.~\eqref{eq:shift_kx_gauge}. \\
\begin{figure}
    \centering
  \includegraphics[width=0.99\linewidth]{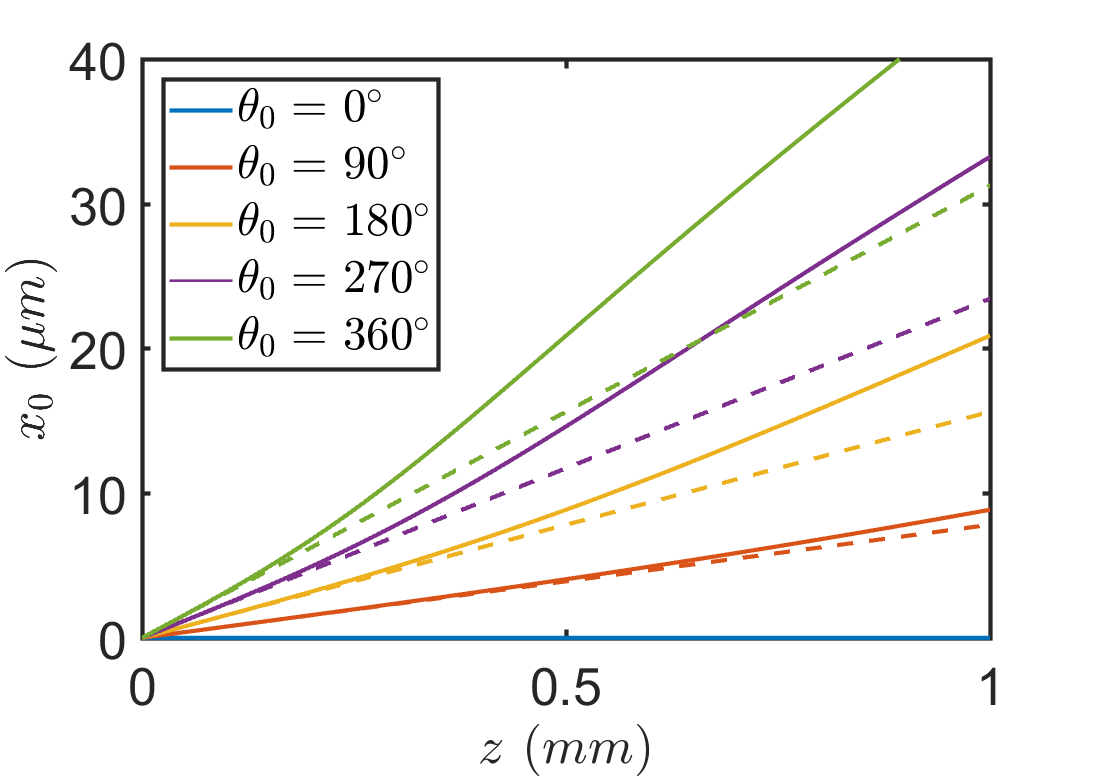}
    \caption{ Trajectories computed via Eq.~\eqref{eq:trajectory_waves} (solid lines) and via Eq.~\eqref{eq:trajectory} (dashed lines) for different values of $\theta_0$. Here $\lambda=1~\mu$m, $n_\bot=1.5$, $n_\|=1.7$, and $L=20~\mu$m.}
    \label{fig:traj_kapitza}
\end{figure}
In the case of inhomogeneous $\bm A$, the beam is deflected more than the theoretical prediction for large $\theta_0$ (see Fig.~\ref{fig:CP_vs_theta0}): evidently, such a difference cannot be ascribed to the linear approximation for $\theta$, which should indeed predict a larger overall shift than the real one. The discrepancy comes from the fact that, until now, we neglected the role of the scalar potential $\phi$, whose weight is actually quadratic with $\theta_0$. Such a potential applies an additional force to the wave, which tends to push the beam towards positive $x$, that is, to enhance the SHE. Quantitatively speaking, the potential $\phi$ adds a new term to the Ehrenfest's theorem determining the beam trajectory, now reading
\begin{multline}
     \label{eq:trajectory_waves}
    \frac{d^2 x_0}{dz^2} =   \pm \frac{ \Delta n}{\overline{n}} \left.\frac{\partial\theta}{\partial x}\right|_{x_0} \sin\left({k_0\Delta n z}\right)  \\ \pm \frac{2}{k_0 \overline{n}} \left. \frac{dx_0}{dz} \frac{\partial^2 \theta}{\partial x^2}\right|_{x_0} \sin^2\left( \frac{k_0 \Delta n z}{2} \right)  
    - \left. \frac{4}{k_0^2 \overline{n}^2}\frac{\partial \theta}{\partial x} \frac{\partial^2 \theta}{\partial x^2}\right|_{x_0},
\end{multline}
bundled with the initial condition $dx_0/dz=0$. 
The first two terms on the RHS of Eq.~\eqref{eq:trajectory_waves} comes from the derivation with respect to $z$ of Eq.~\eqref{eq:velocity},  whereas the last term is the gradient of the Kapitza potential. Figure~\ref{fig:traj_kapitza} shows the trajectories computed from Eq.~\eqref{eq:trajectory_waves} versus $\theta_0$ in a given material. When compared to the approximated trajectories given by Eq.~\eqref{eq:trajectory}, the Kapitza effect pushes the beam away from the center $x=0$, in fact altering the average of the transverse wavevector $k_x$, as witnessed by the changes in the slope observed even at large $z$. \\
\begin{figure}
    \centering
  \includegraphics[width=0.99\linewidth]{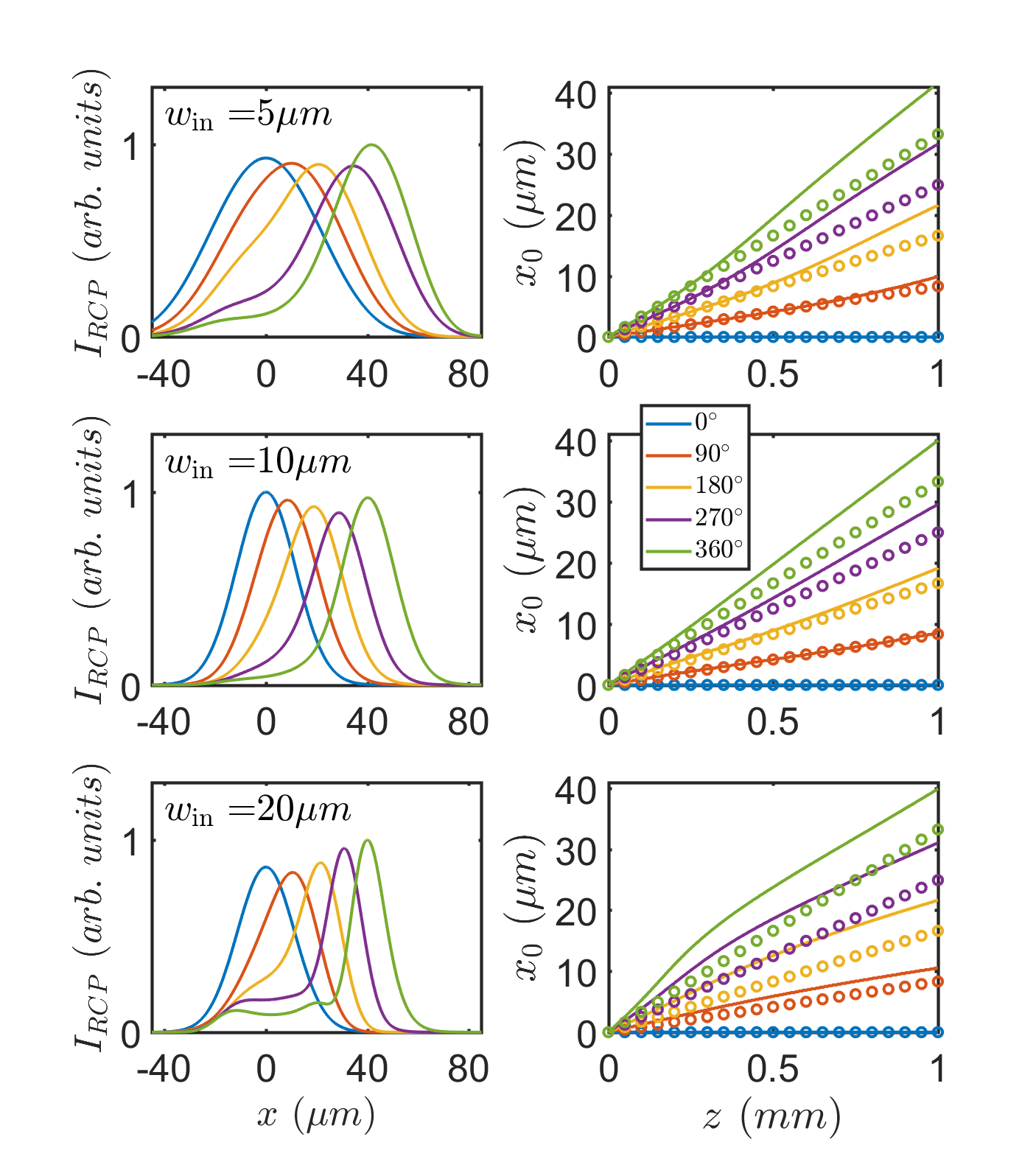}
    \caption{ Intensity profile of the RCP components at its maximum closest to $z=1~$mm (left column) and the corresponding trajectory in the plane $xz$ (right column). Each row corresponds to a different value of the input beam width $w_\mathrm{in}$, reported on the left panels. Each curve corresponds to a different value of $\theta_0$, provided by the legend in the middle row. Circles are predictions from Eq.~\eqref{eq:trajectory}. Here RCP is launched at the input, $\lambda=1064~$nm, and $L=20~\mu$m.}
    \label{fig:different_widths}
\end{figure}
To check this idea, we simulated the beam behavior with the BPM for different input beam widths $w_\mathrm{in}$, thus modifying the spatial overlap of the beam with the distribution of $\theta(x)$. Results for the output profile and beam trajectories are shown in Fig.~\ref{fig:different_widths}. Regardless of the beam size, the shift is larger than what is predicted by Eq.~\eqref{eq:trajectory}, the difference increasing for larger $\theta_0$. The trajectory depends on the beam width due to the different spatial overlap with the gradient in the geometric phase, proportional to $\partial_x \theta$. For large $\theta_0$ and narrow enough beams, the deviation from the ray-optics prediction is small at short $z$, and suddenly increases when $x_0\approx 15\mu$m, corresponding to the point where $\partial^2_x \theta$ reaches its maximum, in agreement with Eq.~\eqref{eq:trajectory_waves}. The beam profiles show how the potential $\bm A$ strongly affects the shape of the beam. For $w_\mathrm{in}=5~\mu$m, the diffraction cone is wide enough to have a substantial coupling of light with the left side of the barrier. Accordingly, the deformation decreases with increasing $\theta_0$ because the light perceives an additional force  due to the Kapitza effect directed towards positive $x$. For $w_\mathrm{in}=10~\mu$m, the leak of power towards negative $x$ is smaller due to the narrower diffraction cone. For $w_\mathrm{in}=20~\mu$m, the beam is broad enough at the input to substantially overlap with the left side of the barrier. This portion of the energy is then subject to a SHE on the opposite direction. In agreement with our interpretation, the leaked energy does not change abruptly with $\theta_0$, and a local minimum is observed between the two peaks of the beam.
Finally, in accordance with Fig.~\ref{fig:traj_kapitza}, the final slope of the beam differs from the ray-optics prediction, the divergence being greater for narrower beams, somehow  recalling the so-called angular Goos-H\"{a}nchen shift occurring at interfaces between different materials \cite{Bliokh:2013}.

\section{Conclusions}

In this work we have presented a new type of PSHE occurring in bulk twisted anisotropic materials capable of achieving transverse shifts of several wavelengths and angular deflections on the range of degrees. We showed how the trajectory of circularly polarized optical beams split into two mirror-symmetric according to their input helicity. We explained the effect in terms of a gauge field associated with the point-wise rotation of the principal axes of the anisotropic material. Both paraxial simulations and full numerical simulations based upon FDTD confirms the existence of this effect, eventually providing a good reciprocal agreement as well. We also discussed how this spin-dependent deflection depends on the anisotropy of the material, proving how the average beam trajectory is almost independent from the material birefringence, if the latter overcomes a given threshold value. 

With respect to future works, on the theoretical side generalization to the full 3D case -thus including the role of orbital angular momentum \cite{Marrucci:2006}- is currently underway; on the experimental side, we are planning to investigate the effect using laser-written nanogratings \cite{Drevinskas:2017}. 
Finally, all along the Paper we emphasized the strong resemblances between optical propagation and matter waves propagating in the presence of a twisted magnetic field: this novel PSHE could represent a new way to realize spin filters for both elemental and composite particles \cite{Karimi:2012}. Further computations including the full Lorentz force can then be used to determine if the proposed configuration is capable of polarizing a beam of electrons using a gradient in the magnetic field \cite{Mott:1929,Batelaan:1997,Kohda:2012}.   

\section{Acknowledgments}

C.P.J. work has been funded by the EU via the H2020 Marie Skłodowska-Curie Actions (fellowship number 889525). We acknowledge the financial support of Deutsche Forschungsgemeinschaft (DFG) through the Collaborative Research Center CRC 1375-NOA (Nonlinear Optics down to Atomic scale).

\appendix

\section{Computation of the trajectory via ray-optics}
\label{app:trajectory}

\begin{figure}
    \centering
  \includegraphics[width=0.99\linewidth]{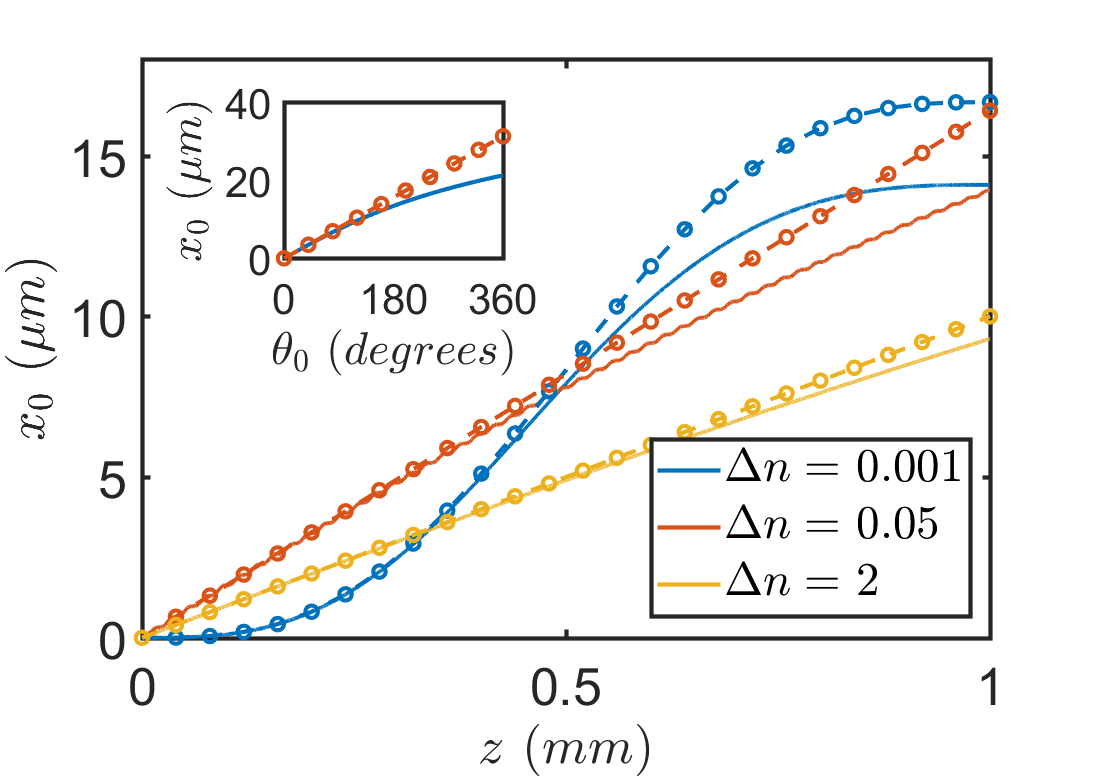}
    \caption{Beam trajectories $x_0$ versus $z$ for three different values of birefringence $\Delta n$ when $n_\bot=1.5$, $\lambda=1064~$nm, and $\theta_0=180^\circ$. Solid lines and dashed lines with circles are computed using Eq.~\eqref{eq:traj_exact} and Eq.~\eqref{eq:trajectory}, respectively. Inset: output position taken in $z=1~$mm versus $\theta_0$ when $\Delta n=0.2$.}
    \label{fig:traj_tanh}
\end{figure}

The wavefront evolution allows us the computation of the local slope of the trajectory as follows
\begin{equation}  \label{eq:velocity}
    \frac{d x_0}{dz} =   \pm \frac{2}{k_0 \overline{n}} \frac{\partial\theta(x_0)}{\partial x} \sin^2\left(\frac{k_0\Delta n z}{2}\right).
\end{equation}
Taking $x_0(z=0)=0$, direct integration provides
\begin{equation} \label{eq:traj_exact}
    x_0(z) =  \pm \frac{2}{k_0 \overline{n}} \int_0^z{ \frac{\partial\theta(x_0)}{\partial x} \sin^2\left(\frac{k_0\Delta n z^\prime}{2}\right) dz^\prime}.
\end{equation}
For large enough birefringence we find
\begin{equation}
    x_0(z) \approx  \pm \frac{1}{k_0 \overline{n}} \int_0^z{ \frac{\partial\theta(x_0)}{\partial x} dz^\prime}.
\end{equation}
Last equation is an integral equation, which needs to be solved numerically in the general case. Figure~\ref{fig:traj_tanh} compares the exact solutions provided by Eq.~\eqref{eq:traj_exact} with the analytical expression~\eqref{eq:trajectory} computed assuming a twisting angle linear in $x$, with a slope equal to the hyperbolic tangent case in the origin $x=0$. The two solutions are similar up to the point where the beam moves far enough from the center, where indeed the first order Taylor expansion for the hyberbolic tangent ceases to be valid. Accordingly, the differences are larger for small $\Delta n$ and when the overall rotation angle increases (see the inset).

\section{Details on the derivation of the paraxial model}
\label{app:derivation_paraxial_model}
Substitution back into Eq.~\eqref{eq:spinorial_E} yields \cite{Jisha:2017_1,Jisha:2017}
\begin{multline}
   \frac{\partial^2 \bm{\psi^\prime}}{\partial z^2}  + k_0^2 \bm{\epsilon}_D \bm{\psi^\prime}     = \\
 -   \frac{\partial^2 \bm{\psi^\prime}}{\partial x^2} + \left(\frac{\partial\theta}{\partial x} \right)^2  \bm{\psi^\prime}   + i \frac{\partial^2 \theta}{\partial x^2}  \bm{\sigma_2}\cdot \bm{\psi^\prime} + 2i  \frac{\partial\theta}{\partial x} \bm{\sigma_2} \cdot \frac{\partial \bm{\psi^\prime}}{\partial  x}.  \label{eq:maxwell_rotated_inho}
\end{multline}
In the next step, we need to transform Eq.~\eqref{eq:maxwell_rotated_inho} into a first order equation after applying the paraxial approximation \cite{Jisha:2017}. To eliminate the dielectric tensor $\bm \epsilon_D$ in Eq.~\eqref{eq:maxwell_rotated_inho}, we find inspiration from the interaction picture in quantum mechanics \cite{Slussarenko:2016}. We indeed use the transformation $\bm \psi^\prime = e^{ik_0\bm N z} \cdot \bm u$ with $\bm N = \left(n_\bot,0;0,n_\| \right)$ \cite{Slussarenko:2016}; in terms of Pauli matrices, $e^{ik_0\bm N z}=e^{ik_0\overline{n}z} e^{-i\bm \sigma_3  \frac{k_0\Delta n z}{2}}$. This transformation factors out the polarization variation due to the material birefringence, thus acting as the unperturbed Hamiltonian in quantum mechanics: in other words, the field $\bm u$ does not change in propagation when the material is homogeneous. The Pauli matrices $\bm \sigma_2$ and $\bm \sigma_3$  do not commute, thus the two last terms in Eq.~\eqref{eq:maxwell_rotated_inho} change their form after the transformation. Using the standard algebra of Pauli matrices, we find $\bm{\tilde{\sigma}}_2 (z) = \bm \sigma_2 \cos(k_0\Delta n z) + \bm \sigma_1 \sin(k_0\Delta n z)$. 

Let us now discuss briefly the numerical method implemented to solve the paraxial vectorial equation. Eq.~\eqref{eq:maxwell_rotated_inho_paraxial} is now a first-order diffusion-like equation, which can be solved assuming unidirectional propagation as in a standard BPM. Numerically, we implemented an operator splitting, where the Crank-Nicolson (CN) scheme is used to solve the diffraction operator. The term proportional to $\partial_x \theta$ is included in the CN algorithm opportunely modified, whereas the term proportional to $\partial_x^2 \theta$ is solved together with the scalar potential using an exponential matrix.

\section{Transformation to the circular basis}
\label{app:transformation_to_circular}
To pass from the linear to the circular basis, we use the transformation $\bm \psi_{LR} = \bm{P} \cdot \bm \psi$, where $\bm P=\left(1/\sqrt{2} \right)\left(1,-i;1, i  \right)$. In terms of Pauli matrices $\bm P = 0.5 e^{-i\pi/4}\left(\bm \sigma_1 + \bm \sigma_2 + \bm \sigma_3 + i\bm I \right)$ and $\bm P^{-1} = 0.5 e^{i\pi/4}\left(\bm \sigma_1 + \bm \sigma_2 + \bm \sigma_3 - i\bm I \right)$, being $\bm I$ the $2\times 2$ identity matrix.
We finally find the new expression in the CP (or LR) basis for the matrix operators in Eq.~\eqref{eq:spinorial_E} by using the identities $\bm P \cdot \bm \sigma_1 \cdot \bm P^{-1}=\bm \sigma_2$ and $\bm P \cdot \bm \sigma_3 \cdot \bm P^{-1}=\bm \sigma_1$. 

\section{Behavior of the polarization over one birefringence length}

\begin{figure}
    \centering
  \includegraphics[width=0.99\linewidth]{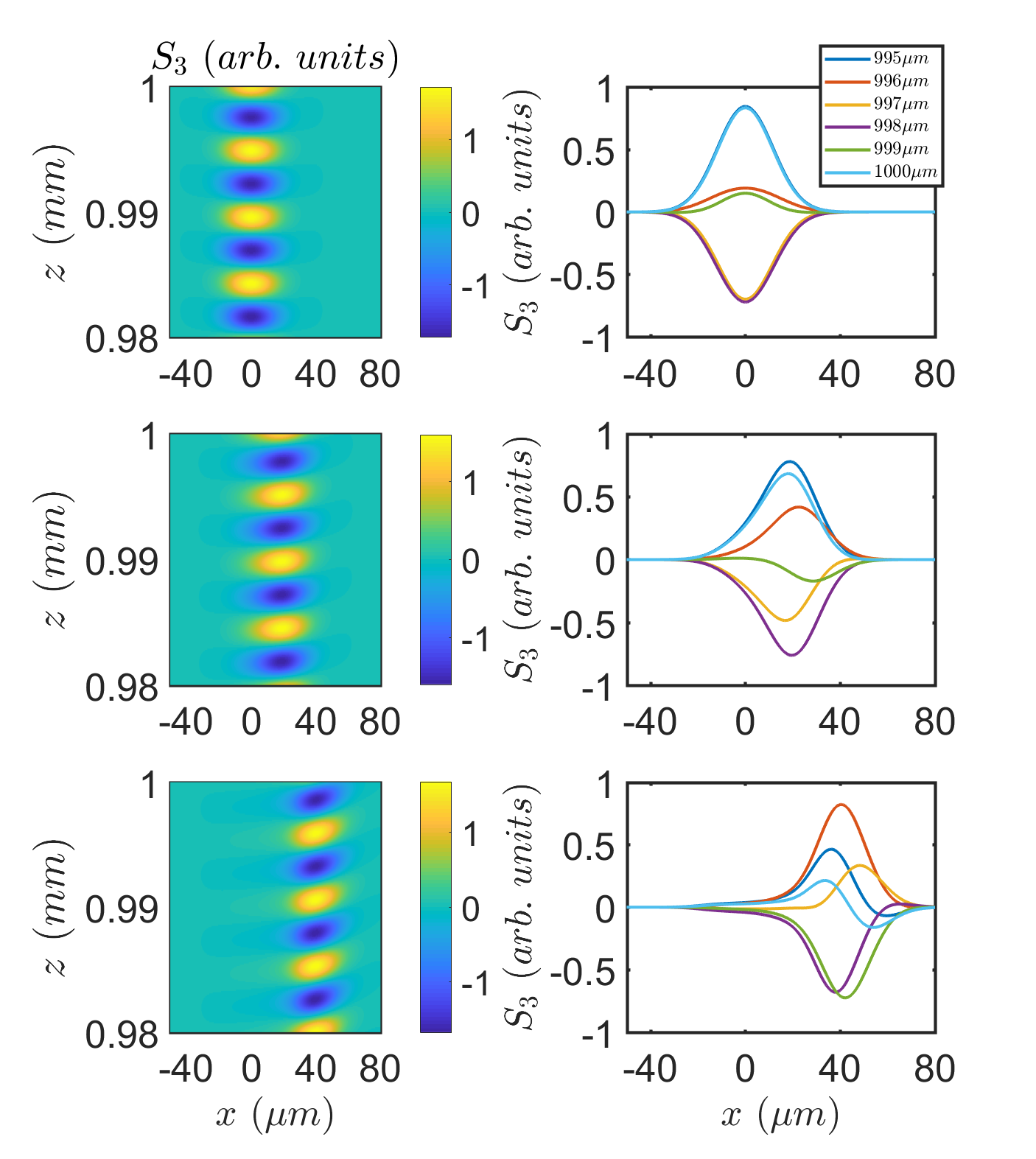}
    \caption{Left column: Stokes parameter $S_3$ plotted around $z=1~$mm for an interval of four birefringence lengths well inside the material. Right column: corresponding five cross-sections cut across $z$, selected equally spaced into a single birefringence length (see the legend).}
    \label{fig:S3_xz}
\end{figure}

Figure~\ref{fig:S3_xz} shows the distribution of the Stokes parameter $S_3$ in the interval $0.98~$mm$<z<1~$mm, corresponding thus to approximately four birefringence length given that $\Delta n=0.2$ and $\lambda=1064~$nm. The profile of $\theta$ is hyperbolic tangent, with $L=20~\mu$m and $\theta_0=0,\ 180^\circ$ and $360^\circ$ from top to bottom. The PSHE induces a tilt in the wavevector, in turn inducing a rotation on the distribution of the Stokes parameter. Despite the inhomogeneous twisting angle, $S_3$ oscillates periodically in propagation with a period given by $\lambda/\Delta n$. 

More details on the transverse profile of $S_3$ can be evinced from the right column of Fig.~\ref{fig:S3_xz}. The cross-section demonstrates  that the beam profile develops a long tail towards negative $x$ for large enough $\theta_0$. Furthermore, for small $\theta_0$ the transverse profile is always bell-shaped, whereas for large $\theta_0$ negative and positive values for $S_3$ are observed in a fixed plane $z=const$. This can be mainly ascribed to the strong tilt induced in the averaged wavevector. 

\section{Behavior versus the birefringence material in the rotated framework}

\label{app:vs_birefringence}
\begin{figure}
    \centering
  \includegraphics[width=0.99\linewidth]{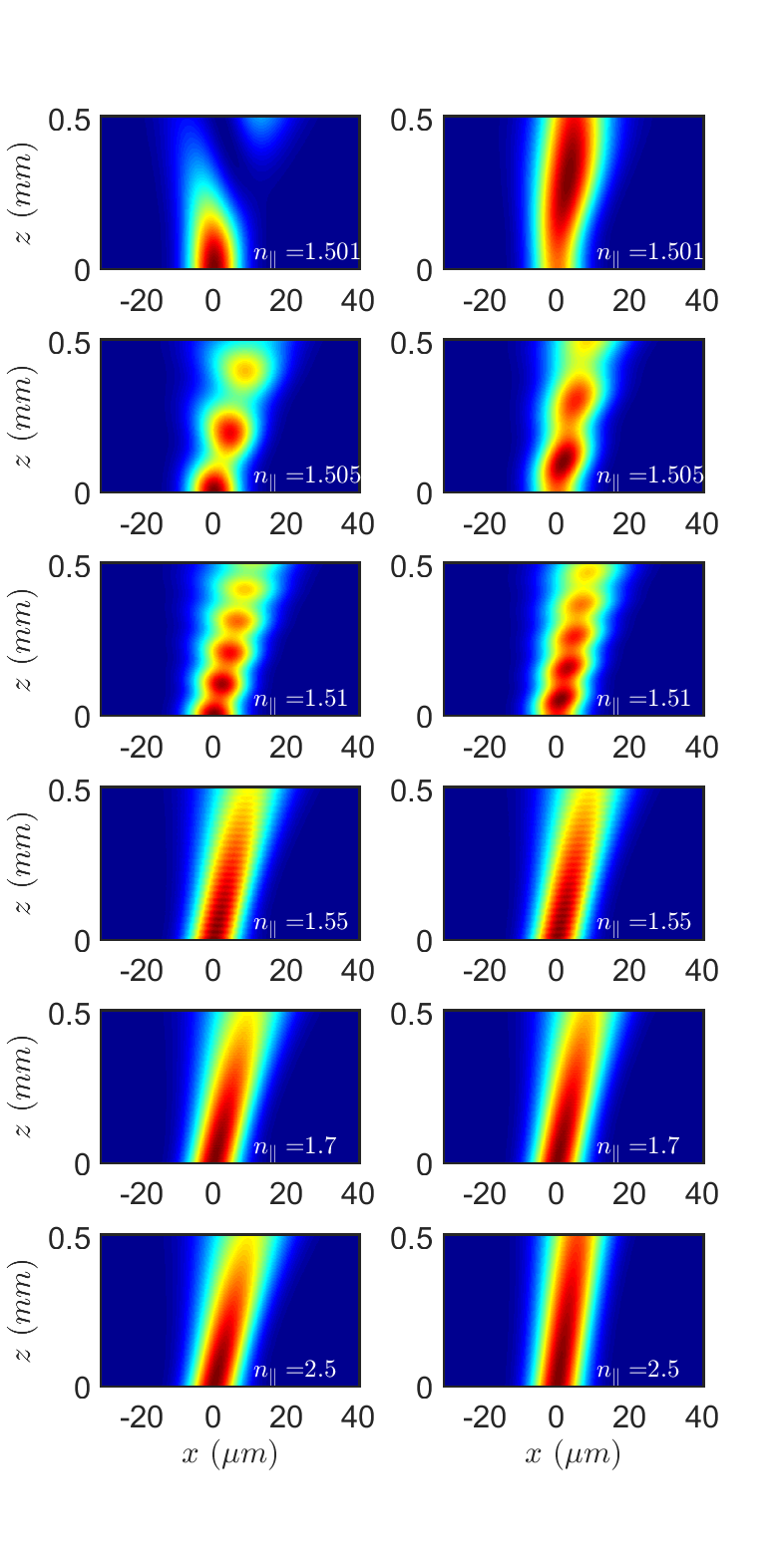}
    \caption{ Plot of $|u(1)|^2$ (left column, the local ordinary component) and $|u(2)|^2$ (right column, the local extraordinary component) on the plane $xz$ for six different values of $n_\|$ (see label in each panel). Wavelength is 1064~nm, $ \theta_0=180^\circ$, $w_\mathrm{in}=10~\mu$m and $L=20\mu$m.}
    \label{fig:profiles_rotated_system}
\end{figure}

Figure~\ref{fig:profiles_rotated_system} shows the intensity profile for the field computed in the rotated framework using Eq.~\eqref{eq:maxwell_rotated_inho_paraxial}. For very low birefringence ($\Delta n=0.001$), diffraction is dominating into a single birefringent length, and the spatial distribution of the two polarizations are very different. Already for $\Delta n=0.005$ the two polarizations almost overlap during propagation, but the intensity is longitudinally modulated due to the diffractive spreading along one birefringence length. At $\Delta n=0.01$ the two profiles are now similar, with a small residual longitudinal periodic modulation in the intensity. At $\Delta n=0.05$ the two profiles are nearly identical, and the fields in the rotated basis are quasi-homogeneous along the propagation distance $z$. At very large birefringence ($\Delta n=1$), the component locally parallel to the extraordinary axis starts to spread less than the other component and to slightly bend towards the $z$ axis. As a matter of fact, Eq.~\eqref{eq:propagation_SVEA_pauli_form} shows that the extraordinary perceives a Kapitza potential barrier lower than the ordinary component by a factor $\frac{1-\gamma}{1+\gamma}$, thus explaining the smaller repulsion with respect to the center $x=0$. Noteworthy, the profile of the other component remains practically unvaried with respect to the lower anisotropy case.


\bibliography{apssamp}

\end{document}